\documentclass[12pt,reqno]{amsart}

\usepackage{epsfig}
\usepackage{amsmath, amsthm, amsfonts}
\usepackage{graphicx}

\numberwithin{equation}{section}

\newcommand{\R}{{\mathbb R}}
\newcommand{\C}{{\mathbb C}}

\renewcommand{\Re}{{\operatorname{Re\,}}}
\renewcommand{\Im}{{\operatorname{Im\,}}}

\newcommand{\Tr}{{{\operatorname{Tr}}}}

\newcommand{\diag}{{\operatorname{diag}}}

\newcommand{\Ai}{{\operatorname{Ai}}}
\newcommand{\al}{\alpha}
\newcommand{\be}{\beta}

\newcommand{\Ga}{\Gamma}
\newcommand{\La}{\Lambda}
\newcommand{\la}{\lambda}

\newcommand{\f}{\varphi}

\newcommand{\om}{\omega}
\newcommand{\Om}{\Omega}

\newtheorem{theo}{{\sc \bf Theorem}}[section]

\newtheorem{prop}[theo]{{\sc \bf Proposition}}

\newenvironment{rem}{\medskip\noindent{\it Remark:\/} }{\medskip}

\sc
\title[Random matrices with external source]
{Large $n$ limit of Gaussian random matrices with external source, part I}
\author{Pavel M. Bleher}
\address{Department of Mathematical Sciences, Indiana
University-Purdue University Indianapolis,
402 N. Blackford St., Indianapolis, IN 46202, U.S.A.}
\email{bleher@math.iupui.edu}
\author{Arno B.J. Kuijlaars}
\address{Department of Mathematics,
Katholieke Universiteit Leuven,
Celestijnenlaan 200 B,
B-3001 Leuven
BELGIUM}
\email{arno@wis.kuleuven.ac.be}

\date{\today}

\dedicatory{Dedicated to Freeman Dyson on his eightieth birthday}
\thanks{The first author was supported in part by
NSF Grant DMS-9970625. The second author was supported by projects
G.0176.02 and G.0455.04 of FWO-Flanders, and INTAS project
03-51-6637.}

\rm

\begin{document}

\begin{abstract}
We consider the random matrix ensemble with an external source
\[ \frac{1}{Z_n} e^{-n \Tr(\frac{1}{2}M^2 -AM)} dM \]
defined on $n\times n$ Hermitian matrices, where $A$ is a diagonal
matrix with only two eigenvalues $\pm a$ of equal multiplicity.
For the case $a > 1$, we establish the universal behavior of local
eigenvalue correlations in the limit $n \to \infty$, which is known
from unitarily invariant random matrix models. Thus, local eigenvalue
correlations are expressed in terms of the sine kernel in the bulk
and in terms of the Airy kernel at the edge of the spectrum.
We use a characterization of the associated multiple Hermite polynomials
by a $3 \times 3$-matrix Riemann-Hilbert problem, and the Deift/Zhou
steepest descent method to analyze the Riemann-Hilbert problem in the
large $n$ limit.
\end{abstract}

\maketitle

\section{Introduction and statement of results}
We will consider the  random matrix ensemble with an external source,
\begin{equation}\label{m1}
\mu_n(dM)=\frac{1}{Z_n}\,e^{-n\Tr(V(M)-AM)}dM,
\end{equation}
defined on $n\times n$ Hermitian matrices $M$. The number $n$ is a large
parameter in the ensemble. The Gaussian ensemble, $V(M)= \frac{1}{2} M^2$,
has been solved in the papers of Pastur \cite{Pas} and Br\'ezin-Hikami \cite{BH1}--\cite{BH5},
by using spectral methods and a contour integration formula
for the determinantal kernel.
In the present work we will develop a completely different approach
to the solution of the Gaussian ensemble with external source. Our approach is
based on the Riemann-Hilbert problem and it is applicable,
in principle, to a general $V$.

We will assume that the external source $A$ is a fixed
diagonal matrix with $n_1$ eigenvalues $a$ and $n_2$ eigenvalues $(-a)$,
\begin{equation}\label{m2}
A=\diag(\underbrace{a,\ldots,a}_{n_1},\underbrace{-a,\ldots,-a}_{n_2}),
\quad n_1+n_2=n.
\end{equation}
As shown by P.~Zinn-Justin \cite{ZJ1}, for any $m\ge 1$, the $m$-point correlation  function
of eigenvalues of $M$ has the determinantal form,
\begin{equation}\label{m3}
R_m(\la_1,\ldots,\la_m)=\det(K_n(\la_j,\la_k))_{1\le j,k\le m}.
\end{equation}
In our previous work \cite{BK} we show that
the kernel $K_n(x,y)$ can be expressed
in terms of a solution to the
following matrix Riemann-Hilbert (RH) problem: find $Y:\C\setminus
\R\to\C^{3\times 3}$ such that
\begin{itemize}
\item $Y$ is analytic on $\C\setminus\R$,
\item for $x\in\R$, we have
\begin{equation}\label{m4}
Y_+(x)=Y_-(x)
\begin{pmatrix}
1 & w_1(x) & w_2(x) \\
0 & 1 & 0 \\
0 & 0 & 1
\end{pmatrix},
\end{equation}
where
\begin{equation}\label{m5}
w_1(x)=e^{-n(V(x)-ax)},\qquad
w_2(x)=e^{-n(V(x)+ax)},
\end{equation}
and $Y_+(x)$ ($Y_-(x)$) denotes the limit of $Y(z)$ as $z\to x$ from
the upper (lower) half-plane,
\item as $z\to\infty$, we have
\begin{equation}\label{m6}
Y(z)=\left( I+O\left(\frac{1}{z}\right)\right)
\begin{pmatrix}
z^n & 0 & 0 \\
0 & z^{-n_1} & 0 \\
0 & 0 & z^{-n_2}
\end{pmatrix},
\end{equation}
where $I$ denotes the $3\times 3$ identity matrix.
\end{itemize}
Namely,
\begin{equation}\label{m7}
\begin{aligned}
K_n(x,y)& =e^{-\frac{1}{2}n(V(x)+V(y))}\frac
{e^{nay}[Y(y)^{-1}Y(x)]_{21}+e^{-nay}[Y(y)^{-1}Y(x)]_{31}}
{2\pi i(x-y)}.
\\
& = \frac{e^{-\frac{1}{2} n (V(x) + V(y))}}{2\pi i(x-y)}
  \begin{pmatrix} 0 & e^{nay} & e^{-nay}\end{pmatrix} Y(y)^{-1} Y(x)
    \begin{pmatrix} 1 \\ 0 \\ 0 \end{pmatrix}.
\end{aligned}
\end{equation}
The RH problem has a unique solution and the solution is expressed
in terms of multiple orthogonal polynomials, see \cite{BK} and
Section 2.1 below. For now, let us mention that the $(1,1)$ entry
$Y_{11}$ satisfies
\begin{align} \label{Y11}
    Y_{11}(z) = \mathbb E \left[ \det( zI - M)\right]
\end{align}
where $\mathbb E$ denotes expectation with respect to the
measure (\ref{m1}). So it is the average characteristic polynomial
for the random matrix ensemble.

It is the aim of this paper to analyze the RH problem as $n \to \infty$, by
using the method of steepest descent / stationary phase of Deift and Zhou
\cite{DZ1}. We focus here on the Gaussian case $V(x) = \frac{1}{2} x^2$.
Our first  result concerns the limiting mean eigenvalue density.
\begin{theo} \label{maintheo1}
Let $V(M) = \frac{1}{2} M^2$, $n_1 = n_2 = n/2$ (so $n$ is even)
and let $a > 1$. Then the limiting mean density of eigenvalues
\begin{equation} \label{rho1}
 \rho(x) = \lim_{n \to \infty} \frac{1}{n} K_n(x,x)
\end{equation}
exists, and it is supported by two intervals, $[-z_1,-z_2]$
and  $[z_2,z_1]$. The density $\rho(x)$
is expressed as
\begin{equation} \label{pastur1}
    \rho(x)=\frac{1}{\pi} \left|\, \Im\xi(x)\right|,
\end{equation}
where $\xi=\xi(x)$ solve the cubic equation,
\begin{equation} \label{pastur2}
    \xi^3-x\xi^2-(a^2-1)\xi+xa^2=0
\end{equation}
(Pastur's equation).
The density $\rho$ is real analytic on $(-z_1,-z_2) \cup (z_2,z_1)$
and it vanishes like a square root at the edge points of its support, i.e.,
there exist constants $\rho_1, \rho_2 > 0$ such that
\begin{equation} \label{rho2}
\begin{aligned}
    \rho(x) & = \frac{\rho_j}{\pi} |x - z_j|^{1/2} (1 + o(1))
        \quad \text{\rm as } x \to z_j, \, x \in (z_2,z_1), \\
    \rho(x) & = \frac{\rho_j}{\pi} |x + z_j|^{1/2} (1 + o(1))
        \quad \text{\rm as } x \to -z_j, \, x \in (-z_1, -z_2).
\end{aligned}
\end{equation}
\end{theo}

\begin{rem}
We obtain $\rho$ from an analysis of the equation
\begin{equation} \label{rem1}
    z = \frac{\xi^3 - (a^2-1) \xi}{\xi^2 - a^2}.
\end{equation}
The critical points of the mapping (\ref{rem1}) satisfy
\begin{equation} \label{rem2}
    \xi^2 = \frac{1}{2} + a^2 \pm \frac{1}{2}\sqrt{1+8a^2}.
\end{equation}
For $a > 1$, the four critical points are real, and they correspond
to four real branch points $\pm z_1$, $\pm z_2$ with $z_1 > z_2 > 0$.
We denote the three inverses of (\ref{rem1}) by $\xi_j(z)$, $j=1,2,3$, where
$\xi_1$ is chosen such that $\xi_1(z) \sim z$ as $z \to \infty$. Then
$\xi_1$ has an analytic continuation to $\C \setminus ([-z_1, -z_2] \cup [z_2, z_1])$
and $\Im \xi_{1+}(x) > 0$ for $x \in (-z_1,-z_2) \cup (z_2,z_1)$. Then
the density $\rho$ is
\begin{equation} \label{rem3}
    \rho(x) = \frac{1}{\pi} \Im \xi_{1+}(x),
\end{equation}
see Section 3.

The assumption $a > 1$ is essential for four real branch points
and a limiting mean eigenvalue density which is supported on two
disjoint intervals.
For $0 < a < 1$, two branch points are purely imaginary, and the limiting
mean eigenvalue density is supported on one interval. The main theorem
on the local eigenvalue correlations continues to hold, but its proof requires
a different analysis of the RH problem. This will be done in part II.
In part III we will discuss the case $a =1$.
\end{rem}

\begin{rem}
The density $\rho$ can also be characterized by a minimization
problem for logarithmic potentials.
Consider the following energy functional defined on pairs $(\mu_1, \mu_2)$
of measures:
\[
\begin{aligned}
E(\mu_1, \mu_2) & = \iint \log \frac{1}{|x-y|} d\mu_1(x) d\mu_1(x)
    + \iint \log \frac{1}{|x-y|} d\mu_2(x) d\mu_2(y) \\
 {}  & + \iint \log \frac{1}{|x-y|} d\mu_1(x) d\mu_2(y)
    + \int \left(\frac{1}{2} x^2 - ax\right) d\mu_1(x)
    + \int \left(\frac{1}{2} x^2 + ax\right) d\mu_2(x).
\end{aligned} \]
The problem is to minimize $E(\mu_1,\mu_2)$ among all pairs $(\mu_1, \mu_2)$ of measures
on $\R$ with $\int d\mu_1 = \int d\mu_2 = \frac{1}{2}$. There is a unique
minimizer, and for $a > 1$, it can be shown that $\mu_1$ is supported on $[z_2,z_1]$,
$\mu_2$ is supported on $[-z_1,-z_2]$ and $\rho$ is the density of $\mu_1 + \mu_2$.
This minimal energy problem is similar to the minimal energy problem
for Angelesco systems in the theory of multiple orthogonal polynomials,
see \cite{AS,GR}.

It is possible to base the asymptotic analysis of the RH problem on the
minimization problem, as done by Deift et al, see \cite{Deift,DKMVZ1,DKMVZ2}, for
the unitarily invariant random matrix model. However, we will not
pursue that here.
\end{rem}

Our main results concern the universality of local eigenvalue correlations
in the large $n$ limit. This was established for unitarily invariant random matrix models
\begin{equation} \label{univ0}
    \frac{1}{Z_n} e^{-n \Tr V(M)} dM
\end{equation}
by Bleher and Its \cite{BI1} for a quartic polynomial $V$, and by Deift et al
\cite{DKMVZ1} for general real analytic $V$.
The universality may be expressed by the following limit
\begin{equation} \label{univ1}
    \lim_{n\to\infty} \frac{1}{n \rho(x_0)} K_n \left(x_0 + \frac{u}{n\rho(x_0)},
    x_0 + \frac{v}{n\rho(x_0)} \right) = \frac{\sin \pi(u-v)}{\pi(u-v)}
\end{equation}
which is valid for $x_0$ in the bulk of the spectrum, i.e., for $x_0$ such that
the limiting mean eigenvalue density $\rho(x_0)$ is positive. The proof of (\ref{univ1})
established Dyson's universality conjecture \cite{Dys,Meh} for unitary ensembles.

In our case, we use a rescaled version of the kernel $K_n$
\begin{equation} \label{m16}
\hat{K}_n(x,y) = e^{n(h(x)-h(y))} K_n(x,y)
\end{equation}
for some function $h$. The rescaling (\ref{m16}) is allowed because it
does not affect the correlation functions $R_m$ (\ref{m3}), which are
expressed as determinants of the kernel.
Note that the kernel $K_n$ of (\ref{m7}) is non-symmetric and there is no
obvious a priori scaling for it.
The function $h$ in (\ref{m16}) has the following form on $(-z_1, -z_2) \cup (z_2,z_1)$
\begin{equation} \label{m20}
    h(x) = - \frac{1}{4} x^2 + \Re \la_{1+}(x), \qquad x \in (-z_1, -z_2) \cup (z_2,z_1)
\end{equation}
with $\la_{1+}(x) = \int_{z_1}^x \xi_{1+}(s) ds$, where $\xi_1$ is as in the first
remark after Theorem \ref{maintheo1}.

\begin{theo} \label{maintheo2}
Let $V(M) = \frac{1}{2} M^2$, $n_1 = n_2 = n/2$, and let $a > 1$. Let $z_1, z_2$
and $\rho$ be as in Theorem {\rm \ref{maintheo1}} and let $\hat{K}_n$ be as in
{\rm (\ref{m16})}. Then for every
$x_0 \in (-z_1, -z_2) \cup (z_2,z_1)$ and $u,v \in \R$, we have
\begin{equation} \label{m17}
    \lim_{n\to\infty}
    \frac{1}{n\rho(x_0)}
    \hat{K}_n \left(x_0 + \frac{u}{n\rho(x_0)}, x_0 + \frac{v}{n \rho(x_0)}\right)
    = \frac{\sin  \pi(u-v)}{\pi(u-v)}.
    \end{equation}
\end{theo}

Our final result concerns the eigenvalue correlations near the edge points
$\pm z_j$. For unitarily invariant random matrix ensembles (\ref{univ0}) the
local correlations near edge points are expressed
in the limit  $n \to \infty$ in terms of the Airy kernel
\begin{equation} \label{univ3}
    \frac{\Ai(u) \Ai'(v) - \Ai'(u) \Ai(v)}{u-v},
\end{equation}
provided that the limiting mean eigenvalue density vanishes like a square root,
which is generically the case \cite{KMc}. In our non-unitarily invariant random matrix
model, the limiting mean eigenvalue density vanishes like a square root, (\ref{rho2}),
and indeed we recover the kernel (\ref{univ3}) in the limit $n\to\infty$.

\begin{theo} \label{maintheo3}
We use the same notation as in Theorem {\rm \ref{maintheo2}}.
Let $\rho_1$ and $\rho_2$ be the constants from {\rm (\ref{rho2})}. Then
for every $u, v \in \R$ we have
\begin{equation} \label{m18}
\lim_{n \to \infty}
    \frac{1}{(\rho_1 n)^{2/3}} \hat{K}_n\left(z_1 + \frac{u}{(\rho_1 n)^{2/3}}, z_1 + \frac{v}{(\rho_1 n)^{2/3}}\right)
    = \frac{\Ai(u)\Ai'(v)-\Ai'(u)\Ai(v)}{u-v},
    \end{equation}
and
\begin{equation} \label{m19}
\lim_{n \to \infty}
    \frac{1}{(\rho_2 n)^{2/3}} \hat{K}_n\left(z_2 - \frac{u}{(\rho_2 n)^{2/3}}, z_2 - \frac{v}{(\rho_2 n)^{2/3}}\right)
    = \frac{\Ai(u)\Ai'(v)-\Ai'(u)\Ai(v)}{u-v}.
    \end{equation}
Similar limits hold near the edge points $-z_1$ and $-z_2$.
\end{theo}

As said before, our results follow from an asymptotic analysis of
the RH problem (\ref{m4})--(\ref{m6}), which involves $3\times 3$
matrices. In the past, asymptotics for RH problems has mostly been
restricted to $2\times 2$-matrix valued RH problems, see e.g.\
\cite{BI1,BI2,DKMVZ1,DKMVZ2} and references cited therein.
The first asymptotic analysis of a $3\times 3$ matrix RH problem
appeared in \cite{KVAW} in an approximation problem for the exponential
function. In the present work we use  some of the ideas of \cite{KVAW}.

As in \cite{KVAW} a main tool in the analysis is an appropriate three
sheeted Riemann surface. To motivate the choice of the Riemann surface
we describe in Section \ref{sec2} the recurrence relations and
differential equations that are satisfied by a matrix $\Psi$, which
is an easy modification of $Y$, see (\ref{m13}) below. The Riemann
surface is studied in Section \ref{sec3} and we obtain from it
the functions $\xi_j$ and $\la_j$, $j=1,2,3$, that are necessary
for the transformations of the RH problem. The first transformation
$Y \mapsto T$ normalizes the RH problem at infinity and at the same
time introduces oscillating diagonal entries in the jump matrices on the cuts
$[-z_1,-z_2]$ and $[z_2,z_1]$, see Section \ref{sec4}. The second transformation $T \mapsto S$
involves opening of lenses around the cuts, which results in
a RH problem for $S$ with rapidly decaying off-diagonal entries
in the jump matrices on the upper and lower boundaries of the lenses,
see Section \ref{sec5}.
The next step is the construction of a parametrix, an approximate
solution to the RH problem. In Section \ref{sec6} we ignore all jumps in
the RH problem for $S$, except those on the cuts $[-z_1,-z_2]$, $[z_2,z_1]$.
This leads to a model RH problem, which we solve by lifting it to
the Riemann surface via the functions $\xi_k$. This leads to the
parametrix away from the edge points $\pm z_1, \pm z_2$. A separate
construction is needed near the edge points. This is done in
Section \ref{sec7} where we build the local parametrices out
of Airy functions. The final transformation $S \mapsto R$ is
done in Section \ref{sec8} and it leads to a RH problem for
$R$ whose jump matrices are uniformly close to the identity matrix.
Then we can use estimates on solution of RH problems, see \cite{Deift},
to conclude that $R$ is close to the identiy matrix, with error
estimates. Having that we can give the proofs of the theorems
in Section \ref{sec9}.

Our approach proves simultaneously large $n$ asymptotics of
the $(1,1)$ entry of $Y$, which by (\ref{Y11}) is equal to the
average characteristic polynomial. This polynomial is
called a multiple Hermite polynomial for the case of
$V(x) = \frac{1}{2}x^2$, see \cite{BK} and Section
\ref{sec2} below.
Since its asymptotics may be of independent interest,
we consider it briefly in Section \ref{sec10} below.
More information on multiple orthogonal polynomials
and their asymptotics can be found in \cite{GR,NS,Nut},
see also the surveys \cite{Apt,AS} and the references
cited therein.

\section{Recurrence relations and differential equations}
\label{sec2}

In order to motivate the introduction of the Riemann surface associated with
(\ref{rem1}) we discuss here
the recurrence relations and differential equations that are satisfied by the
solution of the Riemann-Hilbert problem (\ref{m4})--(\ref{m6})
in case $V(x) = \frac{1}{2}x^2$. It also reveals the integrable structure.
We note however, that the results of this subsection are not essential
for the rest of the paper.

For the recurrence relations we need to separate the indices $n_1$
and $n_2$ in the asymptotic behavior (\ref{m6}) from the exponent
$n$ in the weight functions $w_1$, $w_2$ of (\ref{m5}). In this section we
put
\begin{equation} \label{m5a}
    w_1(x) = e^{-N(\frac{1}{2} x^2-ax)}, \qquad w_2(x) = e^{-N(\frac{1}{2}x^2+ax)}
\end{equation}
where $N$ is fixed, and we let $Y = Y_{n_1,n_2}$ be the solution of the Riemann-Hilbert problem
(\ref{m4}), (\ref{m6}) with $V(x) = \frac{1}{2}x^2$ and $w_1$, $w_2$ given
by (\ref{m5a}).
Let $P_{n_1,n_2}(x)=x^n+\cdots$ be a monic
polynomial of degree $n = n_1+n_2$ such that for $k=1,2$,
\begin{equation}\label{m8}
\int_{-\infty}^\infty P_{n_1,n_2}(x)x^j w_k(x)dx=0, \qquad 0\le j\le n_k-1.
\end{equation}
The polynomial $P_{n_1,n_2}(x)$ is unique and it is called a
multiple Hermite polynomial, see \cite{ABVA,VAC}. Denote for $k=1,2$,
\begin{equation}\label{m9}
h_{n_1,n_2}^{(k)}=\int_{-\infty}^\infty P_{n_1,n_2}(x)x^{n_k}w_k(x)dx\not=0.
\end{equation}
The solution to the RH problem is
\begin{equation}\label{m10}
Y_{n_1,n_2} =
\begin{pmatrix}
P_{n_1,n_2} & C(P_{n_1,n_2}w_1) & C(P_{n_1,n_2}w_2) \\
c_1P_{n_1-1,n_2} & c_1C(P_{n_1-1,n_2}w_1) & c_1C(P_{n_1-1,n_2}w_2) \\
c_2P_{n_1,n_2-1} & c_2C(P_{n_1,n_2-1}w_1) & c_2C(P_{n_1,n_2-1}w_2)
\end{pmatrix},
\end{equation}
with the constants
\begin{equation}\label{m11}
c_1=-\frac{2\pi i}{h_{n_1-1,n_2}^{(1)}}\,,\quad
c_2=-\frac{2\pi i}{h_{n_1,n_2-1}^{(2)}}\,,
\end{equation}
and where $Cf$ denotes the Cauchy transform of $f$,
\begin{equation}\label{m12}
Cf(z)=\frac{1}{2\pi i}\int_{\R}\frac{f(s)}{s-z}\,ds.
\end{equation}

The recurrence relations and differential
equations are nicer formulated in terms of the function
\begin{equation}\label{m13}
\begin{aligned}
\Psi_{n_1,n_2}&=
\begin{pmatrix}
P_{n_1,n_2}e^{-\frac{1}{2}N z^2} & C(P_{n_1,n_2}w_1)e^{-Naz} & C(P_{n_1,n_2}w_2)e^{Naz} \\
P_{n_1-1,n_2}e^{-\frac{1}{2}N z^2} & C(P_{n_1-1,n_2}w_1)e^{-Naz} &
C(P_{n_1-1,n_2}w_2)e^{Naz} \\
P_{n_1,n_2-1}e^{-\frac{1}{2}N z^2} & C(P_{n_1,n_2-1}w_1)e^{-Naz} &
C(P_{n_1,n_2-1}w_2)e^{Naz}
\end{pmatrix}\\
{}&=
\begin{pmatrix}
1 & 0 & 0 \\
0 & c_1^{-1} & 0 \\
0 & 0 & c_2^{-1}
\end{pmatrix} Y_{n_1,n_2}
\begin{pmatrix}
e^{-\frac{1}{2}N z^2} & 0 & 0 \\
0 & e^{-Naz} & 0 \\
0 & 0 & e^{Naz}
\end{pmatrix}.
\end{aligned}
\end{equation}

The function $\Psi = \Psi_{n_1,n_2}$ solves the following RH problem:
\begin{itemize}
\item $\Psi$ is analytic on $\C\setminus\R$,
\item for $x\in\R$, we have
\begin{equation}\label{m14}
\Psi_+(x)=\Psi_-(x)
\begin{pmatrix}
1 & 1 & 1 \\
0 & 1 & 0 \\
0 & 0 & 1
\end{pmatrix},
\end{equation}
\item as $z\to\infty$, we have
\begin{equation}\label{m15}
\Psi(z)=\left( I+O\left(\frac{1}{z}\right)\right)
\begin{pmatrix}
z^n e^{-\frac{1}{2}Nz^2} & 0 & 0 \\
0 & c_1^{-1}z^{-n_1}e^{-Naz} & 0 \\
0 & 0 & c_2^{-1}z^{-n_2}e^{Naz}
\end{pmatrix}.
\end{equation}
\end{itemize}

\begin{prop}\label{Lax} We have the recurrence relations,
\begin{equation}\label{Lax1}
\begin{aligned}
\Psi_{n_1+1,n_2}(z)&=\begin{pmatrix}
z-a & -\frac{n_1}{N}
& -\frac{n_2}{N}  \\
1 & 0 & 0 \\
1 & 0 & -2a
\end{pmatrix}\Psi_{n_1,n_2}(z),\\
\Psi_{n_1,n_2+1}(z)&=\begin{pmatrix}
z+a & -\frac{n_1}{N}
& -\frac{n_2}{N}  \\
1 & 2a & 0 \\
1 & 0 & 0
\end{pmatrix}\Psi_{n_1,n_2}(z),
\end{aligned}
\end{equation}
and the differential equation,
\begin{equation}\label{Lax2}
\Psi_{n_1,n_2}'(z)=N \begin{pmatrix}
-z
 & \frac{n_1}{N} & \frac{n_2}{N}  \\
{}-1 & -a & 0 \\
{}-1 & 0 & a
\end{pmatrix}\Psi_{n_1,n_2}(z).
\end{equation}
\end{prop}

The proof of Proposition \ref{Lax} is given in the Appendix \ref{AI} below.

\bigskip

We look for a WKB solution of the differential equation (\ref{Lax2})
of the form
\begin{equation}\label{wkb1}
\Psi_{n_1,n_2}(z) =W(z)e^{-N\La(z)},
\end{equation}
where $\La$ is a diagonal matrix. By substituting
this form into (\ref{Lax2}) we obtain the equation,
\begin{equation}\label{wkb2}
-W\La'W^{-1}
=A-\frac{1}{N}
W'W^{-1},
\end{equation}
where $A$ is the matrix of coefficients in (\ref{Lax2}).
By dropping the last term we reduce it to the eigenvalue problem,
\begin{equation}\label{wkb3}
W\La'W^{-1}= -A.
\end{equation}
The characteristic polynomial is
\begin{equation}\label{wkb4}
\begin{aligned}
\det \left[ \xi I+A\right]
&=\left|\begin{matrix}
\xi-z
 & t_1 & t_2  \\
{}-1 & \xi-a & 0 \\
{}-1 & 0 & \xi+a
\end{matrix}\right|\\
&=\xi^3-z\xi^2+(t_1+t_2-a^2)\xi+(t_1-t_2+za)a,
\end{aligned}
\end{equation}
where $t_1=\frac{n_1}{N}$ and $t_2=\frac{n_2}{N}$.

The spectral curve
$\xi^3-z\xi^2+(t_1+t_2-a^2)\xi+(t_1-t_2+za)a =0$ defines
a Riemann surface, which in the case of interest in this paper
(where $N = n$ and $n_1 = n_2 = \frac{1}{2}n$) reduces to
\begin{equation} \label{bp0}
    \xi^3 - z \xi^2 - (a^2-1) \xi + z a^2 = 0.
\end{equation}
This defines the Riemann surface that will play a central
role in the rest of the paper.

\section{Riemann surface} \label{sec3}

The Riemann surface is given by the equation (\ref{bp0})
or, if we solve for $z$,
\begin{equation} \label{bp1}
  z = \frac{\xi^3 - (a^2-1) \xi}{\xi^2-a^2}.
\end{equation}
There are three inverse functions to (\ref{bp1}), which
we choose such that as $z \to \infty$,
\begin{equation}\label{wkb5}
\begin{aligned}
\xi_1(z)&=z -\frac{1}{z}+ O\left(\frac{1}{z^3}\right), \\
\xi_2(z)&=a +\frac{1}{2z} +O\left(\frac{1}{z^2}\right), \\
\xi_3(z)&=-a +\frac{1}{2z}+ O\left(\frac{1}{z^2}\right).
\end{aligned}
\end{equation}

We need to find the sheet structure of the Riemann surface
(\ref{bp0}). The critical points of $z(\xi)$ satisfy the equation
\begin{equation}\label{bp4}
\xi^4-(1+2a^2)\xi^2+(a^2-1)a^2=0,
\end{equation}
which is biquadratic. The roots are
\begin{equation}\label{bp5}
\xi^2_{1,2}=\frac{1}{2}+a^2\pm\frac{1}{2}\sqrt{1+8a^2}.
\end{equation}
The value $a=1$ is critical, in the sense that for $a>1$ all the roots
are real, while for $a< 1$, two are real and two are purely imaginary.
In this paper we will consider the case $a>1$. As noted before, we will
consider the cases $a<1$ and $a=1$ in parts II and III.

Set
\begin{equation}\label{bp6}
p,q=\sqrt{\frac{1}{2}+a^2\mp\frac{1}{2}\sqrt{1+8a^2}},
\qquad 0<p<q.
\end{equation}
Then the critical points are $\xi=\pm p,\;\pm q$. The branch points
on the $z$-plane are $\pm z_1$ and $\pm z_2$,  where
\begin{equation}\label{bp7}
z_1=q\frac{\sqrt{1+8a^2}+ 3}
{\sqrt{1+8a^2}+ 1}\,,\quad
z_2=p\frac{\sqrt{1+8a^2}- 3}
{\sqrt{1+8a^2}- 1},\qquad 0<z_2<z_1.
\end{equation}

We can show that $\xi_1$, $\xi_2$, and $\xi_3$ have analytic
extensions to $\mathbb C \setminus
([-z_1,-z_2] \cup [z_2,z_1])$, $\mathbb C \setminus [z_2,z_1]$ and
$\mathbb C \setminus [-z_1, -z_2]$, respectively.
Also on the cut $[z_2,z_1]$,
\begin{equation}\label{bp8}
\begin{aligned}
\xi_{1+}(x)=\overline{\xi_{1-}(x)}=\xi_{2-}(x)=\overline{\xi_{2+}(x)}, &
\quad z_2\le x\le z_1, \\
\Im \xi_{1+}(x) > 0, \qquad \qquad & \quad z_2 < x < z_1,
\end{aligned}
\end{equation}
 and $\xi_3(x)$ is real. On the cut $[-z_1,-z_2]$,
\begin{equation}\label{bp9}
\begin{aligned}
\xi_{1+}(x)=\overline{\xi_{1-}(x)}=\xi_{3-}(x)=\overline{\xi_{3+}(x)}, &
\quad -z_1\le x\le -z_2, \\
\Im \xi_{1+}(x) > 0, \qquad \qquad & \quad -z_1 < x < -z_2,
\end{aligned}
\end{equation}
and $\xi_2(x)$ is real. Figure 1 depicts the three sheets of the
Riemann surface (\ref{bp0}).

\begin{figure}
\scalebox{1.0}{\includegraphics{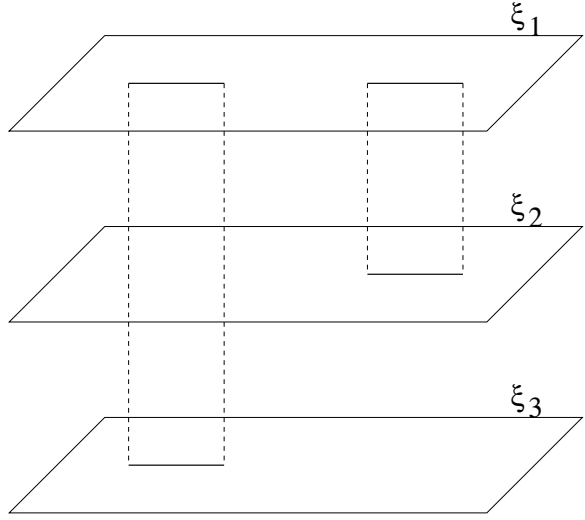}}
\caption{The three sheets of the Riemann surface}
\label{fig1}
\end{figure}

We define
\begin{equation} \label{defrho}
\rho(x) = \frac{1}{\pi} \Im \xi_{1+}(x),
    \qquad x \in [-z_1,-z_2] \cup [z_2, z_1].
\end{equation}

\begin{prop} \label{rho-density}
We have
$\rho(x) > 0$ for $x \in (-z_1, -z_2) \cup (z_2, z_1)$
and
\begin{equation} \label{rhoint}
    \int_{-z_1}^{-z_2} \rho(x) dx = \int_{z_2}^{z_1} \rho(x) dx = \frac{1}{2}.
\end{equation}
Moreover, there are $\rho_1, \rho_2 > 0$ such that
\begin{equation} \label{rhoconst}
\begin{aligned}
    \rho(x) & = \frac{\rho_j}{\pi} |x - z_j|^{1/2} \left(1 + O(x-z_j)\right)
        \quad \text{\rm as } x \to z_j, \, x \in (z_2,z_1) \\
    \rho(x) & = \frac{\rho_j}{\pi} |x + z_j|^{1/2} \left(1 + O(x+z_j)\right)
        \quad \text{\rm as } x \to -z_j, \, x \in (-z_1, -z_2).
\end{aligned}
\end{equation}
\end{prop}

\textbf{Proof}:
The fact that $\rho(x) > 0$ for $x \in (-z_1,-z_2) \cup (z_2,z_1)$
was already noted in (\ref{bp8}) and (\ref{bp9}).

We have for $x \in [z_2,z_1]$,
\[ \rho(x) = \frac{1}{2\pi} \Im (\xi_{1+}(x) - \xi_{1-}(x))
= \frac{1}{2\pi i} (\xi_{1+}(x) - \xi_{1-}(x)) =
\frac{1}{2\pi i} (\xi_{2-}(x) - \xi_{2+}(x)). \]
Thus
\[ \int_{z_2}^{z_1} \rho(x) dx = \frac{1}{2\pi i} \oint_{\gamma} \xi_2(z) dz \]
where $\gamma$ is a contour encircling the interval $[z_2,z_1]$ once
in the positive direction. Letting the contour go to infinity and
using the asymptotic behavior (\ref{wkb5}) of $\xi_2(z)$ as $z \to \infty$,
we find the value of the second integral in (\ref{rhoint}). The value of
the first integral follows in the same way.

For (\ref{rhoconst}) we note that near the branch point $z_1$, we have
for a constant $\rho_1 > 0$,
\begin{equation} \label{rhoconst2}
\begin{aligned}
    \xi_1(z) &= q +  \rho_1 (z-z_1)^{1/2} + O\left(z-z_1\right) \\
    \xi_2(z) &= q -  \rho_1 (z-z_1)^{1/2} + O\left(z-z_1\right)
\end{aligned}
\end{equation}
as $z \to z_1$. Similarly, near $z_2$ we have for a constant $\rho_2 > 0$,
\begin{equation} \label{rhoconst3}
\begin{aligned}
    \xi_1(z) &= p -  \rho_2 (z_2 - z)^{1/2} + O\left(z_2-z\right) \\
    \xi_2(z) &= p +  \rho_2 (z_2 - z)^{1/2} + O\left(z_2-z\right)
\end{aligned}
\end{equation}
as $z \to z_2$ (with main branches of the square root). By symmetry, we
have similar expressions near $-z_1$ and $-z_2$ and (\ref{rhoconst})
follows. \hfill \qedsymbol

\bigskip

Next, we need the integrals of the $\xi$-functions,
\begin{equation} \label{wkb5a}
    \la_k(z) = \int^z \xi_k(s) ds, \qquad k =1,2,3,
\end{equation}
which we take so that $\la_1$ and $\la_2$
are analytic on $\mathbb C \setminus (-\infty, z_1]$
and $\la_3$ is analytic on $\mathbb C \setminus (-\infty, -z_2]$.
From (\ref{wkb5}) it follows that, as $z\to \infty$,
\begin{equation}\label{wkb6}
\begin{aligned}
\la_1(z)&=\frac{z^2}{2} - \ln z +l_1 + O\left(\frac{1}{z^2}\right),\\
\la_2(z)&=az +\frac{1}{2}\ln z+l_2+ O\left(\frac{1}{z}\right), \\
\la_3(z)&=-az+\frac{1}{2}\ln z+l_3+O\left(\frac{1}{z}\right),
\end{aligned}
\end{equation}
where $l_1$, $l_2$, $l_3$ are some constants, which we choose as follows.
We choose $l_1$ and $l_2$ such that
\[ \la_1(z_1) = \la_2(z_1) = 0, \]
and then $l_3$ such that
\[ \la_3(-z_2) = \la_{1+}(-z_2) = \la_{1-}(-z_2) - \pi i. \]
Then we have the following jump relations:
\begin{equation} \label{wkb7}
\begin{aligned}
\la_{1+}(x) - \la_{1-}(x) = -\pi i, &
    \qquad x \in [-z_2,z_2], \\
\la_{1+}(x) - \la_{1-}(x) = -2\pi i, &
    \qquad x \in (-\infty, -z_1], \\
\la_{2+}(x) - \la_{2-}(x) = \pi i, &
    \qquad x\in (-\infty, z_2], \\
\la_{1+}(x) = \la_{2-}(x), \quad \la_{1-}(x) = \la_{2+}(x), &
   \qquad x \in [z_2,z_1], \\
\la_{1+}(x) = \la_{3-}(x),  \quad \la_{1-}(x) - \pi  i = \la_{3+}(x), &
    \qquad x \in [-z_1,-z_2]. \\
\la_{3+}(x) - \la_{3-}(x) = \pi i, &
    \qquad x \in (-\infty, -z_1].
\end{aligned}
\end{equation}

Note that due to the first two equations of (\ref{wkb7})
we have that $e^{n\la_1(z)}$ is  analytic
on the complex plane with cuts on $[-z_1,-z_2]$ and
$[z_2,z_1]$ (recall that $n$ is even). Furthermore,
we also see that  $e^{n\la_2(z)}$ (resp., $e^{n\la_3(z)}$)
is  analytic on the complex plane with a cut on $[z_2,z_1]$ (resp.,
$[-z_1,-z_2]$), see Figure \ref{fig1}.

For later use, we state the following two propositions.
\begin{prop} \label{laineq1}
On $\mathbb R \setminus [z_2,z_1]$ we have $\Re \la_{2+} < \Re \la_{1-}$,
and on $\mathbb R \setminus [-z_1,-z_2]$, we have
$\Re \la_{3+} < \Re \la_{1-}$.
\end{prop}
\begin{proof}
It is easy to see that $\xi_1(x) > \xi_2(x)$
for $x > z_1$. Since $\la_1(z_1) = \la_2(z_1)$
and $\la'_j = \xi_j$ for $j=1,2,3$, it is then clear
that $\la_1(x) > \la_2(x)$ for $x > z_1$.

We also have that $\Re \xi_{1-}(x) < \xi_2(x)$ for $x < z_2$,
from which it follows that $\Re \la_{1-}(x) > \Re \la_{2+}(x)$.

Similarly we find that $\Re \la_{3+} < \Re \la_{1-}$ on
$\mathbb R \setminus [-z_1,-z_2]$.
\end{proof}

\begin{prop} \label{laineq2}
\begin{enumerate}
\item[\rm (a)]
The open interval $(z_2,z_1)$ has a neighborhood $U_1$ in the complex
plane such that
\[ \Re \la_3(z) < \Re \la_1(z) < \Re \la_2(z) \]
for every $z \in U_1 \setminus (z_2,z_1)$.
\item[\rm (b)]
The open interval $(-z_1,-z_2)$ has a neighborhood $U_2$ in the complex
plane such that
\[ \Re \la_2(z) < \Re \la_1(z) < \Re \la_3(z) \]
for every $z \in U_2 \setminus (-z_1,-z_2)$.
\end{enumerate}
\end{prop}
\begin{proof}
The function $F = \la_{2+} - \la_{1+}$ is purely imaginary on $(z_2,z_1)$.
Its derivative is $F'(x) = \xi_{2+}(x) - \xi_{1+}(x) = -2\pi i \rho(x)$, and this has
negative imaginary part. The Cauchy Riemann equations then imply that
the real part of $F$ increases as we move from the interval $(z_2,z_1)$ into
the upper half-plane. Thus $\Re \la_2(z) - \Re \la_1(z) > 0$ for $z$ near $(z_2,z_1)$ in the
upper half-plane. Similarly, $\Re \la_2(z) - \Re \la_1(z) > 0$ for $z$
near $(z_2,z_1)$ in the lower half-plane.

By Proposition \ref{laineq1} we have $\Re \la_3 < \Re \la_{1-}$ on $[z_2,z_1]$.
By continuity, the inequality continues to hold in a complex
neighborhood of $[z_2,z_1]$. This proves part (a).
The proof of part (b) is similar.
\end{proof}

\section{First Transformation of the RH Problem}
\label{sec4}

Using the functions $\lambda_j$ and the constants $l_j$, $j=1,2,3$, from the
previous section, we define
\begin{equation} \label{defT}
   T(z) = \diag\left(e^{-nl_1}, e^{-nl_2}, e^{-nl_3}\right)
    Y(z) \diag\left(e^{n(\la_1(z)-\frac{1}{2}z^2)}, e^{n(\la_2(z) - az)}, e^{n(\la_3(z)+az)} \right).
\end{equation}

Then by (\ref{m4}) and (\ref{defT}), we have $T_+(x) = T_-(x) j_T(x)$, $x \in \mathbb R$,
where
\begin{equation}\label{jT}
\begin{aligned}
j_T(x)&=
\begin{pmatrix}
e^{n\left(\la_{1+}(x)-\la_{1-}(x)\right)}
& e^{n\left(\la_{2+}(x)-\la_{1-}(x)\right)}
& e^{n\left(\la_{3+}(x)-\la_{1-}(x)\right)} \\
0 & e^{n\left(\la_{2+}(x)-\la_{2-}(x)\right)} & 0 \\
0 & 0 & e^{n\left(\la_{3+}(x)-\la_{3-}(x)\right)}
\end{pmatrix}.
\end{aligned}
\end{equation}
The jump relations (\ref{wkb7}) allow us to simplify the jump matrix $j_T$
on the different parts of the real axis.
On $[z_2,z_1]$, (\ref{jT}) reduces to
\begin{equation}\label{ft4}
\begin{aligned}
j_T &=\begin{pmatrix}
e^{n(\la_1-\la_2)_+} & 1 & e^{n(\la_{3} - \la_{1-})} \\
0 & e^{n(\la_1-\la_2)_-} & 0 \\
0 & 0 & 1
\end{pmatrix}
\end{aligned}
\end{equation}
and on $[-z_1,-z_2]$ to
\begin{equation}\label{ft7}
\begin{aligned}
j_T&=\begin{pmatrix}
e^{n(\la_1-\la_3)_+} & e^{n(\la_{2+}-\la_{1-})} & 1 \\
0 & 1 & 0 \\
0 & 0 & e^{n(\la_1-\la_3)_-}
\end{pmatrix}.
\end{aligned}
\end{equation}
On $(-\infty,-z_1] \cup [-z_2,z_2]\cup [z_1,\infty)$, (\ref{jT})
reduces to
\begin{equation}\label{ft9}
\begin{aligned}
j_T=\begin{pmatrix}
1
& e^{n\left(\la_{2+}-\la_{1-}\right)}
& e^{n\left(\la_{3+}-\la_{1-}\right)} \\
0 & 1 & 0 \\
0 & 0 & 1
\end{pmatrix},\quad x\in
(-\infty,-z_1] \cup [-z_2,z_2] \cup [z_1,\infty).
\end{aligned}
\end{equation}

The asymptotics of $T$ are, because of (\ref{m6}), (\ref{wkb6}), and (\ref{defT}),
\begin{equation}\label{ft11}
\begin{aligned}
T(z)&= I+O\left(\frac{1}{z}\right) \qquad \mbox{as } z \to \infty.
\end{aligned}
\end{equation}
Thus $T$ solves  the following RH problem:
\begin{itemize}
\item $T$ is analytic on $\C\setminus\R$,
\item
\begin{equation}\label{ft14}
T_+(x)=T_-(x)j_T(x),\quad x\in\R,
\end{equation}
\item as $z\to\infty$,
\begin{equation}\label{ft15}
T(z)=I+O\left(\frac{1}{z}\right).
\end{equation}
\end{itemize}

Using (\ref{defT}) in (\ref{m7}) we see that the kernel $K_n$ can be expressed
in terms of $T$ as follows
\begin{equation} \label{ft16}
    K_n(x,y) = \frac{e^{\frac{1}{4}n(x^2 - y^2)}}{2\pi i(x-y)}
    \begin{pmatrix} 0 & e^{n\la_{2+}(y)} & e^{n\la_{3+}(y)} \end{pmatrix}
    T^{-1}_+(y) T_+(x) \begin{pmatrix} e^{-n\la_{1+}(x)} \\ 0 \\ 0\end{pmatrix}.
\end{equation}

\section{Second Transformation of the RH Problem}
\label{sec5}

\begin{figure}
\scalebox{1.0}{\includegraphics{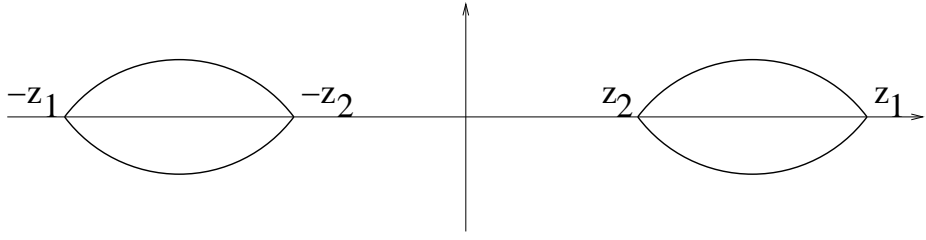}}
\caption{The lenses with vertices $-z_1,-z_2$ and $z_2,z_1$.}
\label{fig2}
\end{figure}

The second transformation of the RH problem is opening of lenses.
Consider a lens with vertices $z_2,z_1$, see Figure \ref{fig2}.
The lens is contained in the neighborhood $U_1$ of $(z_2,z_1)$,
see Proposition \ref{laineq2}.
 We have the factorization,
\begin{equation}\label{st1}
\begin{aligned} { }&
\begin{pmatrix}
e^{n(\la_1-\la_2)_+} & 1 & e^{n(\la_{3+}-\la_{1-})} \\
0 & e^{n(\la_1-\la_2)_-} & 0 \\
0 & 0 & 1
\end{pmatrix} \\
&
=\begin{pmatrix}
1 & 0  & 0 \\
e^{n(\la_1-\la_2)_-} & 1 & -e^{n(\la_3-\la_2)_-} \\
0 & 0 & 1
\end{pmatrix}
\begin{pmatrix}
0 & 1  & 0 \\
-1 & 0 & 0 \\
0 & 0 & 1
\end{pmatrix}
\begin{pmatrix}
1 & 0 & 0\\
e^{n(\la_1-\la_2)_+} & 1 & e^{n(\la_3-\la_2)_+} \\
0 & 0 & 1
\end{pmatrix}.
\end{aligned}
\end{equation}
Set
\begin{equation}\label{st2}
S(z)=\left\{
\begin{aligned}
{}&T(z)\begin{pmatrix}
1 & 0  & 0 \\
-e^{n(\la_1(z)-\la_2(z))} & 1 & -e^{n(\la_3(z)-\la_2(z))} \\
0 & 0 & 1
\end{pmatrix} \;\text{\rm in the upper lens region},\\
{}&T(z)
\begin{pmatrix}
1 & 0 & 0 \\
e^{n(\la_1(z)-\la_2(z))} & 1 & -e^{n(\la_3(z)-\la_2(z))} \\
0 & 0 & 1
\end{pmatrix}\;\text{\rm in the lower lens region}.
\end{aligned}\right.
\end{equation}
Then (\ref{ft14}) and (\ref{st2}) imply that
\begin{equation}\label{st3}
S_+(x)=S_-(x)j_S(x);
\quad j_S(x)=
\begin{pmatrix}
0 & 1 & 0 \\
-1 & 0 & 0 \\
0 & 0 & 1
\end{pmatrix},\quad x\in[z_2,z_1].
\end{equation}
Similarly, consider a lens with vertices $-z_1,-z_2,$
that is  contained in $U_2$ (see Proposition \ref{laineq2})
and set
\begin{equation}\label{st4}
S(z)=\left\{
\begin{aligned}
{}&T(z)\begin{pmatrix}
1 & 0  & 0 \\
0 & 1 & 0 \\
-e^{n(\la_1(z)-\la_3(z))} & -e^{n(\la_2(z)-\la_3(z))} & 1
\end{pmatrix} \;\text{\rm in the upper lens region},\\
{}&T(z)
\begin{pmatrix}
1 & 0 & 0 \\
0 & 1 & 0 \\
e^{n(\la_1(z)-\la_3(z)))} & -e^{n(\la_2(z)-\la_3(z))} & 1
\end{pmatrix}\;\text{\rm in the lower lens region}.
\end{aligned}\right.
\end{equation}
Then (\ref{ft14}) and (\ref{st4}) imply that
\begin{equation}\label{st5}
S_+(x)=S_-(x)j_S(x);\quad
j_S(x)=
\begin{pmatrix}
0 & 0 & 1 \\
0 & 1 & 0 \\
-1 & 0 & 0
\end{pmatrix},\quad x\in[-z_1,-z_2].
\end{equation}
Set
\begin{equation}\label{st6}
S(z)=T(z)\quad\text{\rm outside of the lens regions}.
\end{equation}
Then we have jumps on the boundary of the lenses,
\begin{equation}\label{st7}
S_+(z)=S_-(z)j_S(z),
\end{equation}
where the contours are oriented from left to right
(that is, from $-z_1$ to $-z_2$, or from $z_2$ to $z_1$),
and where $S_+$ ($S_-$) denotes the limiting value of $S$
from the left (right) if we traverse the contour according
to its orientation.
The jump matrix $j_S$ in (\ref{st7}) has the form
\begin{equation}\label{st8}
\begin{aligned}
j_S(z)&=
\begin{pmatrix}
1 & 0 & 0 \\
e^{n(\la_1(z)-\la_2(z))} & 1 & e^{n(\la_3(z)-\la_2(z))} \\
0 & 0 & 1
\end{pmatrix}\;\text{\rm on the upper boundary of
the $[z_2,z_1]$-lens},\\
j_S(z)&=
\begin{pmatrix}
1 & 0 & 0 \\
e^{n(\la_1(z)-\la_2(z))} & 1 & -e^{n(\la_3(z)-\la_2(z))} \\
0 & 0 & 1
\end{pmatrix}\;\text{\rm on the lower boundary of
the $[z_2,z_1]$-lens},\\
j_S(z)&=
\begin{pmatrix}
1 & 0 & 0 \\
0 & 1 & 0 \\
e^{n(\la_1(z)-\la_3(z))} & e^{n(\la_2(z)-\la_3(z))} & 1
\end{pmatrix}\;\text{\rm on the upper boundary of
the $[-z_1,-z_2]$-lens},\\
j_S(z)&=
\begin{pmatrix}
1 & 0 & 0 \\
0 & 1 & 0 \\
e^{n(\la_1(z)-\la_3(z))} & -e^{n(\la_2(z)-\la_3(z))} & 1
\end{pmatrix}\;\text{\rm on the lower boundary of
the $[-z_1,-z_2]$-lens}.
\end{aligned}
\end{equation}
On $(-\infty,z_1]\cup[-z_2,z_2]\cup[z_1,\infty)$, $S$ has the
same jump as $T$, so that
\begin{equation}\label{st9}
S_+(x)=S_-(x)j_S(x);\quad
j_S(x)=j_T(x),\quad x\in (-\infty,z_1]\cup[-z_2,z_2]\cup[z_1,\infty).
\end{equation}
Thus, $S$ solves  the following RH problem:
\begin{itemize}
\item $S$ is analytic on $\C\setminus(\R\cup\Ga)$, where $\Ga$
is the boundary of the lenses,
\item
\begin{equation}\label{st10}
S_+(z)=S_-(z)j_S(z),\quad z\in\R\cup\Ga,
\end{equation}
\item as $z\to\infty$,
\begin{equation}\label{st11}
S(z)=I+O\left(\frac{1}{z}\right).
\end{equation}
\end{itemize}

The kernel $K_n$ is expressed in terms of $S$ as follows,
see (\ref{ft16}) and the definitions (\ref{st2}) and (\ref{st4}).
For $x$ and $y$ in $(z_2,z_1)$ we have
\begin{equation} \label{st12}
K_n(x,y) = \frac{e^{\frac{1}{4}n(x^2-y^2)}}{2\pi i(x-y)}
    \begin{pmatrix} -e^{n\la_{1+}(y)} & e^{n \la_{2+}(y)} & 0 \end{pmatrix}
        S_+^{-1}(y) S_+(x) \begin{pmatrix} e^{-n\la_{1+}(x)} \\ e^{-n\la_{2+}(x)} \\ 0 \end{pmatrix},
\end{equation}
while for $x$ and $y$ in $(-z_1,-z_2)$ we have
\begin{equation} \label{st13}
K_n(x,y) = \frac{e^{\frac{1}{4}n(x^2-y^2)}}{2\pi i(x-y)}
    \begin{pmatrix} -e^{n\la_{1+}(y)} & 0 & e^{n \la_{3+}(y)} \end{pmatrix}
        S_+^{-1}(y) S_+(x) \begin{pmatrix} e^{-n\la_{1+}(x)} \\ 0 \\ e^{-n\la_{3+}(x)} \end{pmatrix}.
\end{equation}

Since $\la_{1+}$ and $\la_{2+}$ are complex conjugates on $(z_2,z_1)$,
we can rewrite (\ref{st12}) for $x,y \in (z_2,z_1)$ as
\begin{equation} \label{st14}
\begin{aligned}
K_n(x,y) & = \frac{e^{n(h(y)-h(x))}}{2\pi i(x-y)}
    \begin{pmatrix} -e^{n i \Im \la_{1+}(y)} & e^{-n i \Im \la_{1+}(y)} & 0 \end{pmatrix}
        S_+^{-1}(y) S_+(x) \begin{pmatrix} e^{-n i \Im \la_{1+}(x)} \\ e^{n i \Im \la_{1+}(x)} \\ 0 \end{pmatrix}
\end{aligned}
\end{equation}
where $h(x) = -\frac{1}{4} x^2 + \Re \la_{1+}(x)$ as in (\ref{m20}).
Similarly, we have for $x,y \in (-z_1,-z_2)$,
\begin{equation} \label{st15}
\begin{aligned}
K_n(x,y) & = \frac{e^{n(h(y)-h(x))}}{2\pi i(x-y)}
    \begin{pmatrix} -e^{n i \Im \la_{1+}(y)} & 0 & e^{-n i \Im \la_{1+}(y)}  \end{pmatrix}
        S_+^{-1}(y) S_+(x) \begin{pmatrix} e^{-n i \Im \la_{1+}(x)} \\ 0 \\ e^{n i \Im \la_{1+}(x)} \\ \end{pmatrix}.
\end{aligned}
\end{equation}

\section{Model RH Problem}
\label{sec6}

As $n\to\infty$, the jump matrix $j_S(z)$ is exponentially close
to the identity matrix at every $z$
outside of $[-z_1,-z_2] \cup [z_2,z_1]$. This follows from (\ref{st8})
and Proposition \ref{laineq2} for $z$ on the boundary of the lenses,
and from (\ref{st9}), (\ref{ft4}) and Proposition \ref{laineq1} for $z$ on the real
intervals $(-\infty, -z_1)$, $(-z_2, z_2)$ and $(z_1, \infty)$.

In this section we solve the following model RH problem, where we ignore
the exponentially small jumps: find $M:\,
\C\setminus ([-z_1,-z_2]\cup[z_2,z_1])\to \C^{3\times 3}$ such that
\begin{itemize}
\item $M$ is analytic on $\C\setminus ([-z_1,-z_2]\cup[z_2,z_1])$,
\item
\begin{equation}\label{mod2}
M_+(x)=M_-(x)j_S(x),\qquad x\in (-z_1,-z_2)\cup (z_2,z_1),
\end{equation}
\item as $z\to\infty$,
\begin{equation}\label{mod3}
M(z)=I+O\left(\frac{1}{z}\right).
\end{equation}
\end{itemize}
This problem is similar to the RH problem considered in \cite[Section 6.1]{KVAW}.
We also follow a similar method to solve it.

\begin{figure}
\scalebox{1.0}{\includegraphics{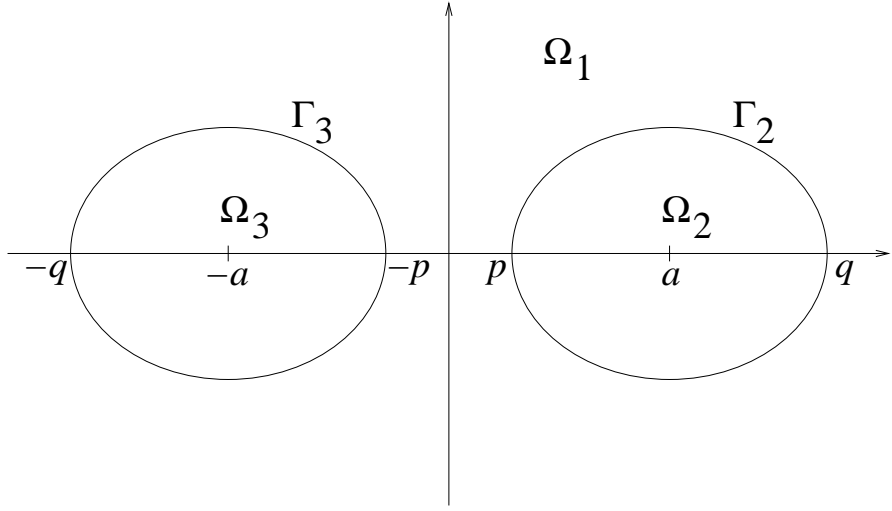}}
\caption{Partition of the complex $\xi$-plane.}\label{fig3}
\end{figure}

We lift the model RH problem to the Riemann surface of
(\ref{bp0}) with the sheet structure as in Figure \ref{fig1}.
Consider to that end the range of the functions $\xi_k$ on the
complex plane,
\begin{equation}\label{mod4}
\begin{aligned}
\Om_1&=\xi_1(\C \setminus ([-z_1,-z_2] \cup [z_2,z_1])),\\
\Om_2&=\xi_2(\C \setminus [z_2,z_1]), \\
\Om_3&=\xi_3(\C \setminus [-z_1,-z_2]).
\end{aligned}
\end{equation}
Then $\Om_1$, $\Om_2$, $\Om_3$ give a partition of the complex
plane into three regions, see Figure \ref{fig3}.
The regions $\Om_2$, $\Om_3$ are bounded,
$a\in\Om_2$, $-a\in\Om_3$, with the symmetry conditions,
\begin{equation}\label{mod5}
\overline{\Om_2}=\Om_2,\quad \overline{\Om_3}=\Om_3,\quad \Om_2=-\Om_3.
\end{equation}
Denote by $\Ga_k$ the boundary of $\Om_k$, $k=2,3$,
see Figure \ref{fig3}. Then we have
\begin{equation}\label{mod6}
\begin{aligned}
\xi_{1+}([z_2,z_1])&=\xi_{2-}([z_2,z_1])
=\Ga_{2}^+\equiv \Ga_2\cap\{\Im z\ge 0\},
\\
\xi_{1-}([z_2,z_1])&=\xi_{2+}([z_2,z_1])
=\Ga_{2}^-\equiv \Ga_2\cap\{\Im z\le 0\},
\\
\xi_{1+}([-z_1,-z_2])&=\xi_{3-}([-z_1,-z_2])
=\Ga_{3}^+\equiv \Ga_3\cap\{\Im z\ge 0\}\\
\xi_{1-}([-z_1,-z_2])&=\xi_{3+}([-z_1,-z_2])
=\Ga_{3}^-\equiv \Ga_3\cap\{\Im z\le 0\}.
\end{aligned}
\end{equation}

We are looking for a solution $M$ in the following form:
\begin{equation}\label{mod7}
M(z)=
\begin{pmatrix}
M_1(\xi_1(z)) & M_1(\xi_2(z)) & M_1(\xi_3(z)) \\
M_2(\xi_1(z)) & M_2(\xi_2(z)) & M_2(\xi_3(z)) \\
M_3(\xi_1(z)) & M_3(\xi_2(z)) & M_3(\xi_3(z))
\end{pmatrix},
\end{equation}
where $M_1(\xi)$, $M_2(\xi)$, $M_3(\xi)$ are three analytic functions
on $\C \setminus (\Ga_1 \cup \Ga_2)$. To satisfy jump condition
(\ref{mod2}) we need the following
relations for $k=1,2,3$:
\begin{equation}\label{mod9}
\begin{aligned}
M_{k+}(\xi)&=M_{k-}(\xi),\quad \xi\in\Ga_{2}^-\cup\Ga_{3}^-, \\
M_{k+}(\xi)&=-M_{k-}(\xi),\quad \xi\in\Ga_{2}^+\cup\Ga_{3}^+.
\end{aligned}
\end{equation}
Since $\xi_1(\infty)=\infty$, $\xi_2(\infty)=a$, $\xi_3(\infty)=-a$,
then to satisfy (\ref{mod3}) we demand
\begin{equation}\label{mod10}
\begin{aligned}
M_1(\infty)&=1,\quad M_1(a)=0,\quad M_1(-a)=0;\\
M_2(\infty)&=0,\quad M_2(a)=1,\quad M_2(-a)=0;\\
M_3(\infty)&=0,\quad M_3(a)=0,\quad M_3(-a)=1.
\end{aligned}
\end{equation}

Equations (\ref{mod9})--(\ref{mod10}) have the following solution:
\begin{equation}\label{mod11}
M_1(\xi)=\frac{\xi^2-a^2}
{\sqrt{(\xi^2-p^2)(\xi^2-q^2)}},\quad
M_{2,3}(\xi)=c_{2,3}\frac{\xi \pm a}
{\sqrt{(\xi^2-p^2)(\xi^2-q^2)}},
\end{equation}
with cuts at $\Ga_{2}^+$, $\Ga_{3}^+$. The constants $c_{2,3}$ are determined
by the equations $M_{2,3}(\pm a)=1$. By (\ref{bp4}),
\begin{equation}\label{mod12}
(\xi^2-p^2)(\xi^2-q^2)
=\xi^4-(1+2a^2)\xi^2+(a^2-1)a^2,
\end{equation}
hence
\begin{equation}\label{mod13}
M_2(a)=c_2\frac{2a}{\sqrt{-2a^2}}.
\end{equation}
By taking into account the cuts of $M_2(\xi)$ we obtain that
\begin{equation}\label{mod13a}
M_2(a)=c_2i\sqrt 2,
\end{equation}
hence
\begin{equation}\label{mod14}
c_2=-\frac{i}{\sqrt 2}.
\end{equation}
Similarly,
\begin{equation}\label{mod15}
M_3(-a)=c_3\frac{-2a}{\sqrt{-2a^2}}=c_3i\sqrt 2,
\end{equation}
hence $c_3$ is the same as $c_2$,
\begin{equation}\label{mod16}
c_3=-\frac{i}{\sqrt 2}.
\end{equation}

Thus, the solution to the model RH problem is given as
\begin{equation}\label{mod17}
M(z)=
\begin{pmatrix}
\frac{\xi_1^2(z)-a^2}
{\sqrt{(\xi_1^2(z)-p^2)(\xi_1^2(z)-q^2)}}
& \frac{\xi_2^2(z)-a^2}
{\sqrt{(\xi_2^2(z)-p^2)(\xi_2^2(z)-q^2)}}
& \frac{\xi_3^2(z)-a^2}
{\sqrt{(\xi_3^2(z)-p^2)(\xi_3^2(z)-q^2)}} \\
-i\frac{\xi_1(z)+a}
{\sqrt{2(\xi_1^2(z)-p^2)(\xi_1^2(z)-q^2)}}
& -i\frac{\xi_2(z)+a}
{\sqrt{2(\xi_2^2(z)-p^2)(\xi_2^2(z)-q^2)}}
& -i\frac{\xi_3(z)+a}
{\sqrt{2(\xi_3^2(z)-p^2)(\xi_3^2(z)-q^2)}} \\
-i\frac{\xi_1(z)-a}
{\sqrt{2(\xi_1^2(z)-p^2)(\xi_1^2(z)-q^2)}}
& -i\frac{\xi_2(z)-a}
{\sqrt{2(\xi_2^2(z)-p^2)(\xi_2^2(z)-q^2)}}
& -i\frac{\xi_3(z)-a}
{\sqrt{2(\xi_3^2(z)-p^2)(\xi_3^2(z)-q^2)}}
\end{pmatrix},
\end{equation}
with cuts on $[z_2,z_1]$ and $[-z_1,-z_2]$.

The model solution $M(z)$ will be used to construct a {\it parametrix}
for the RH problem for $S$ outside of a small neighborhood of the edge
points. Namely, we will fix some $r>0$ and consider the disks
of radius $r$ around the edge points.
At the edge points $M(z)$ is not analytic and in a
neighborhood of the edge points the parametrix is constructed
differently.

\section{Parametrix at  Edge Points}
\label{sec7}

We consider  small disks $D(\pm z_j,r)$ with radius $r > 0$
and centered at the edge points,
and look for a local parametrix $P$ defined on the union of
the four disks such that
\begin{itemize}
\item $P$ is analytic on $D(\pm z_j, r) \setminus(\R\cup\Ga)$,
\item
\begin{equation}\label{lp1}
P_+(z)= P_-(z)j_S(z),\quad z\in (\R\cup\Ga) \cap D(\pm z_j, r),
\end{equation}
\item as $n\to\infty$,
\begin{equation}\label{lp1b}
P(z)=\left(I+O\left(\frac{1}{n}\right) \right) M(z)
\quad \text{\rm uniformly for $z \in \partial D(\pm z_j,r)$}.
\end{equation}
\end{itemize}

We consider here the edge point $z_1$ in detail.
We note that by (\ref{rhoconst}) and (\ref{wkb5a}) we have as $z \to z_1$,
\begin{equation} \label{rhoconst4}
\begin{aligned}
    \la_1(z) &= q(z-z_1) + \frac{2\rho_1}{3} (z-z_1)^{3/2} + O(z-z_1)^{2} \\
    \la_2(z) &= q(z-z_1) - \frac{2\rho_1}{3} (z-z_1)^{3/2} + O(z-z_1)^{2}
\end{aligned}
\end{equation}
so that
\begin{equation} \label{rhoconst5}
    \la_1(z) - \la_2(z) = \frac{4 \rho_1}{3} (z-z_1)^{3/2} + O(z-z_1)^{5/2}
\end{equation}
as $z \to z_1$. Then it follows that
\begin{equation} \label{beta}
\beta(z) = \left[\frac{3}{4}(\la_1(z)-\la_2(z))\right]^{2/3}
\end{equation}
is analytic at $z_1$, real-valued on the real axis near $z_1$
and $\beta'(z_1) = \rho_1^{2/3} > 0$. So $\beta$ is a conformal map from
$D(z_1, r)$ to a convex neighborhood of the origin, if $r$
is sufficiently small (which we assume to be the case).
We take $\Gamma$ near $z_1$ such that
\[ \beta(\Gamma \cap D(z_1,r)) \subset
\{z \mid \arg(z) = \pm 2\pi/3 \}. \]
Then $\Gamma$ and $\R$ divide the disk $D(z_1,r)$ into
four regions numbered I, II, III, and IV, such
that $0 < \arg \beta(z) < 2\pi/3$, $2\pi/3 < \arg \beta(z) < \pi$,
$-\pi < \arg \beta(z) < -2\pi/3$, and $-2\pi/3 < \arg \beta(z) < 0$ for $z$
in regions I, II, III, and IV, respectively.

Recall that the jumps $j_S$ near $z_1$ are given by
(\ref{st3}), (\ref{st8}), and (\ref{ft4}):
\begin{equation} \label{lp2}
\begin{aligned}
j_S&=
\begin{pmatrix} 0 & 1 & 0 \\
-1 & 0 & 0 \\ 0 & 0 & 1 \end{pmatrix}
\qquad \text{\rm on } [z_1-r, z_1) \\
j_S&=
\begin{pmatrix} 1 & 0 & 0 \\
e^{n(\la_1-\la_2)} & 1 & e^{n(\la_3-\la_2)} \\
0 & 0 & 1 \end{pmatrix}
\; \text{\rm on the upper boundary of the lens in $D(z_1,r)$ } \\
j_S&=
\begin{pmatrix} 1 & 0 & 0 \\
e^{n(\la_1-\la_2)} & 1 & -e^{n(\la_3-\la_2)} \\
0 & 0 & 1 \end{pmatrix}
\; \text{\rm on the lower boundary of the lens in $D(z_1,r)$} \\
j_S&=
\begin{pmatrix} 1 & e^{n(\la_2-\la_1)} & e^{n(\la_3-\la_1)} \\
0 & 1 & 0 \\ 0 & 0 & 1 \end{pmatrix}
\qquad \text{\rm on } (z_1, z_1+r].
\end{aligned}
\end{equation}

We write
\begin{equation} \label{tildeP}
\tilde{P} =
\left\{ \begin{array}{ll}
P
\begin{pmatrix}
1 & 0 & 0 \\ 0 & 1 & -e^{n(\la_3-\la_2)} \\
0 & 0 & 1 \end{pmatrix}
& \text{\rm in regions I and IV} \\
P & \text{\rm in regions II and III.}
\end{array} \right.
\end{equation}
Then the jumps for $\tilde{P}$ are
$\tilde{P}_+ = \tilde{P}_- j_{\tilde{P}}$
where
\begin{equation} \label{lp5}
\begin{aligned}
j_{\tilde{P}}&=
\begin{pmatrix} 0 & 1 & 0 \\
-1 & 0 & 0 \\ 0 & 0 & 1 \end{pmatrix}
\qquad \text{\rm on } [z_1-r, z_1) \\
j_{\tilde{P}}&=
\begin{pmatrix} 1 & 0 & 0 \\
e^{n(\la_1-\la_2)} & 1 & 0 \\
0 & 0 & 1 \end{pmatrix}
\; \text{\rm on the upper side of the lens in } D(z_1,r) \\
j_{\tilde{P}}&=
\begin{pmatrix} 1 & 0 & 0 \\
e^{n(\la_1-\la_2)} & 1 & 0 \\
0 & 0 & 1 \end{pmatrix}
\; \text{\rm on the lower side of the lens in } D(z_1,r) \\
j_{\tilde{P}}&=
\begin{pmatrix} 1 & e^{n(\la_2-\la_1)} & 0 \\
0 & 1 & 0 \\ 0 & 0 & 1 \end{pmatrix}
\qquad \text{\rm on } (z_1, z_1+r].
\end{aligned}
\end{equation}
We still have the matching condition
\begin{equation} \label{match2}
\tilde{P}(z) = \left(I+O\left(\frac{1}{n}\right) \right) M(z)
\quad \text{\rm uniformly for $z \in \partial D(z_1,r)$},
\end{equation}
since $\Re \la_3 < \Re \la_2$ on $\overline{D(z_1,r)}$,
which follows from Proposition \ref{laineq1}.

The RH problem for $\tilde{P}$ is essentially a
$2\times 2$ problem, since the jumps (\ref{lp5}) are non-trivial only in
the upper $2\times 2$ block. A solution can be constructed in
a standard way out of Airy functions. The Airy function $\Ai(z)$
solves the equation $y'' = zy$ and for any $\varepsilon >0$, in the
sector $\pi + \varepsilon \leq \arg z \leq \pi - \varepsilon$, it has
the asymptotics as $z \to \infty$,
\begin{equation}\label{ep3}
\Ai(z)=\frac{1}{2\sqrt\pi z^{1/4}}e^{-\frac{2}{3}z^{3/2}}
\left(1+O(z^{-3/2})\right).
\end{equation}
The functions $\Ai(\om z)$, $\Ai(\om^2 z)$, where
$\om=e^{\frac{2\pi i}{3}}$, also solve the equation $y''=zy$, and we
have the linear relation,
\begin{equation}\label{ep4}
\Ai(z)+\om\Ai(\om z)+\om^2\Ai(\om^2 z)=0.
\end{equation}
Write
\begin{equation}\label{ep6}
y_0(z)=\Ai(z), \quad y_1(z)=\om\Ai(\om z),
\quad y_2(z)=\om^2\Ai(\om^2 z),
\end{equation}
and we use these functions to define
\begin{equation}\label{ep5}
\Phi(z)=
\left\{
\begin{aligned}
{}&\begin{pmatrix}
y_0(z) & -y_2(z) & 0 \\
y_0'(z) & -y_2'(z) & 0 \\
0 & 0 & 1
\end{pmatrix},\quad \mbox{for $0 < \arg z < 2\pi/3$},\\
{}&\begin{pmatrix}
-y_1(z) & - y_2(z) & 0 \\
-y_1'(z) & -y_2'(z) & 0 \\
0 & 0 & 1
\end{pmatrix}, \quad \mbox{for $2\pi/3 < \arg z < \pi$}, \\
{}&\begin{pmatrix}
-y_2(z) & y_1(z) & 0 \\
-y_2'(z) & y_1'(z) & 0 \\
0 & 0 & 1
\end{pmatrix}, \quad \mbox{for $-\pi < \arg z < -2\pi/3$}, \\
{}&\begin{pmatrix}
y_0(z) & y_1(z) & 0 \\
y_0'(z) & y_1'(z) & 0 \\
0 & 0 & 1
\end{pmatrix},\quad \mbox{for $-2\pi/3 < \arg z < 0$}.
\end{aligned}\right.
\end{equation}
Then
\begin{equation} \label{ep7}
\tilde{P}(z) = E_n(z) \Phi(n^{2/3} \beta(z))
\diag\left(e^{\frac{1}{2} n (\la_1(z)-\la_2(z))},
e^{-\frac{1}{2}n(\la_1(z) - \la_2(z))}, 1\right)
\end{equation}
where $E_n$ is an analytic prefactor that takes care of the matching
condition (\ref{match2}). Explicitly, $E_n$ is given by
\begin{equation} \label{ep8}
E_n = \sqrt{\pi} M
\begin{pmatrix} 1 & -1 & 0 \\ -i & - i & 0 \\ 0 & 0 & 1
\end{pmatrix}
\begin{pmatrix} n^{1/6} \beta^{1/4} & 0 & 0  \\ 0 & n^{-1/6}
  \beta^{-1/4} & 0 \\ 0 & 0 & 1
\end{pmatrix}.
\end{equation}

A similar construction works for a parametrix $P$
around the other edge points.

\section{Third transformation}
\label{sec8}

In the third and final transformation we put
\begin{equation} \label{tt1}
\begin{aligned}
R(z) & = S(z) M(z)^{-1}
    \quad \text{\rm for $z$ outside the disks $D(\pm z_j, r)$, $j=1,2$} \\
R(z) & = S(z) P(z)^{-1}
    \quad \text{\rm for $z$ inside the disks.}
\end{aligned}
\end{equation}

\begin{figure}
\scalebox{1.0}{\includegraphics{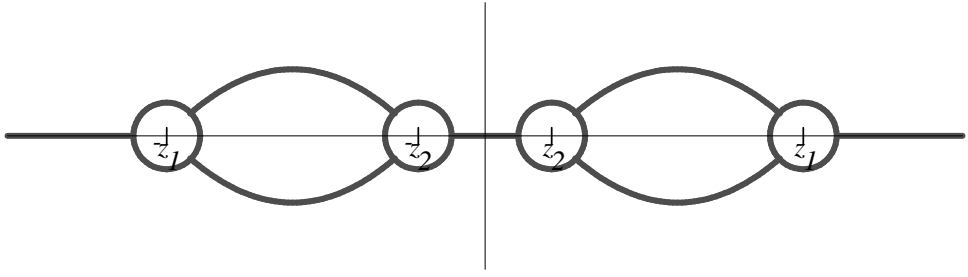}}
\caption{The contour $\Gamma_R$ for $R$.}
\label{fig5}
\end{figure}

Then $R$ is analytic on $\C \setminus \Ga_R$, where $\Ga_R$ consists of
the four circles $\partial D(\pm z_j, r)$, $j=1,2$, the parts of $\Gamma$
outside the four disks, and the real intervals $(-\infty, -z_1-r)$,
$(-z_2+r, z_2-r)$, $(z_1+r,\infty)$, see Figure \ref{fig5}.
There are jump relations
\begin{equation} \label{tt2}
    R_+ = R_- j_R
\end{equation}
where
\begin{equation} \label{tt3}
\begin{aligned}
j_R &= M P^{-1}
    \quad \text{\rm on the circles, oriented counterclockwise} \\
j_R &= M j_s M^{-1}
    \quad \text{\rm on the remaining parts of $\Gamma_R$.}
\end{aligned}
\end{equation}
From (\ref{lp1b}) it follows that $j_R = I +O(1/n)$ uniformly on
the circles, and from (\ref{st8}), (\ref{st9}), (\ref{ft4}) and
Propositions \ref{laineq1} and \ref{laineq2} it follows that
$j_R = I + O(e^{-cn})$ for some $c > 0$ as $n \to \infty$, uniformly on the
remaining parts of $\Gamma_R$. So we can conclude
\begin{equation} \label{tt4}
    j_R(z) = I + O\left(\frac{1}{n}\right)
    \quad \text{\rm as $n \to \infty$, uniformly on $\Gamma_R$.}
\end{equation}
As $z\to \infty$, we have
\begin{equation} \label{tt5}
    R(z) = I + O(1/z).
\end{equation}

From (\ref{tt2}), (\ref{tt4}), (\ref{tt5}) and the fact that we can deform
the contours in any desired direction, it follows that
\begin{equation} \label{tt6}
    R(z) = I + O\left(\frac{1}{n(|z|+1)}\right) \quad \mbox{as } n \to \infty.
\end{equation}
uniformly for $z \in \C \setminus \Ga_R$, see \cite{Deift,DKMVZ1,DKMVZ2,K}.

By Cauchy's theorem, we then also have
\[ R'(z) = O\left(\frac{1}{n(|z|+1)}\right) \]
and thus
\begin{equation} \label{tt7}
\begin{aligned}
    R^{-1}(y) R(x) &= I + R^{-1}(y) \left(R(x) - R(y)\right)
    & = I + O\left(\frac{x-y}{n}\right)
\end{aligned}
\end{equation}
which is the form we will use below.

\section{Proofs of the theorems}
\label{sec9}

\subsection{Proof of Theorem \ref{maintheo1}}
Consider $x \in (z_2,z_1)$.
We may assume that the circles around the edge points are such
that $x$ is outside of the four disks.
Then (\ref{tt1}) shows that $S(x) = R(x)M(x)$ and it follows
easily from (\ref{tt7}) and the fact that $M_+$ is real analytic in a
neighorhood of $x$ that
\begin{equation} \label{pf1}
    S_+^{-1}(y) S_+(x) = I + O\left(x-y\right)
    \qquad \text{\rm as $y \to x$}
\end{equation}
uniformly in $n$.
Thus by (\ref{st14}) we have that
\begin{equation} \label{pf2}
\begin{aligned}
K_n(x,y) &=
  \frac{e^{n(h(y)-h(x))}}{2\pi i  (x-y)}
    \begin{pmatrix} -e^{n i \Im \la_{1+}(y)} &  e^{-n i \Im \la_{1+}(y)} & 0 \end{pmatrix}
       \left( I + O\left(x-y\right)\right) \begin{pmatrix} e^{-n i \Im \la_{1+}(x)} \\  e^{n i \Im \la_{1+}(x)} \\ 0 \end{pmatrix} \\
       & =
       e^{n(h(y)-h(x))} \left[
       \frac{-e^{n i (\Im \la_{1+}(y) - \Im \la_{1+}(x))} + e^{-n i (\Im \la_{1+}(y) - \Im \la_{1+}(x))}}{2\pi i (x-y)}
       +  O(1) \right] \\
       & =
       e^{n(h(y)-h(x))} \left[
       \frac{\sin(n \Im (\la_{1+}(x) - \la_{1+}(y)))}{\pi (x-y)}
       +  O(1) \right] \\
\end{aligned}
\end{equation}
and the $O(1)$ holds uniformly in $n$.
Letting $y \to x$ and noting that by  (\ref{wkb5a}) and (\ref{defrho})
\begin{equation} \label{pf2b}
    \frac{d}{dy} \Im \la_{1+}(y) = \Im \xi_{1+}(y) = \pi \rho(y)
\end{equation}
we obtain by l'Hopital's rule,
\begin{equation} \label{pf3}
K_n(x,x) = n \rho(x) + O(1),
\end{equation}
which proves Theorem \ref{maintheo1} if $x \in (z_2, z_1)$. The proof for
$x \in (-z_1,-z_2)$ is similar, and also follows because of symmetry.

For $x \in (-\infty, -z_1) \cup (-z_2, z_2) \cup (z_1, \infty)$, we have
that $K_n(x,x)$ decreases exponentially fast. For example, for $x > z_1$,
we have that
\begin{equation} \label{pf3b}
    K_n(x,x) = O\left( e^{-n (\la_1(x)-\la_2(x))}\right)
    \quad \text{\rm as } n \to \infty.
\end{equation}
This follows from (\ref{ft16}) and the observation that
that $T_+^{-1}(y) T_+(x) = I + O(x-y)$ as $y \to x$ if $x > z_1$.
It is clear that (\ref{pf3b}) implies
\begin{equation} \label{pf3c}
    \lim_{n \to \infty} \frac{1}{n} K_n(x,x) =0.
\end{equation}

We also have (\ref{pf3c}) if $x$ is one of the edge points.
In fact, for an edge point $x$ it can be shown as in the
proof of Theorem \ref{maintheo3} that
\begin{equation} \label{pf3d}
    \frac{1}{n} K_n(x,x) = O\left(\frac{1}{n^{1/3}}\right)
    \quad \text{\rm as } n \to \infty.
\end{equation}
This completes the proof of Theorem  \ref{maintheo1}.
\hfill \qedsymbol

\subsection{Proof of Theorem \ref{maintheo2}}

We give the proof for $x_0 \in (z_2,z_1)$, the proof for $x_0 \in (-z_1,-z_2)$
being similar. We let
\begin{equation} \label{pf4}
x = x_0 + \frac{u}{n\rho(x_0)}, \qquad y = x_0 + \frac{v}{n\rho(x_0)}.
\end{equation}
Then we have (\ref{pf2}), and so by the definition (\ref{m16}) of $\hat{K}_n$,
\begin{equation} \label{pf5}
\begin{aligned}
\frac{1}{n \rho(x_0)} \hat{K}_n(x,y)
       & =
       \frac{\sin (n \Im (\la_{1+}(x) - \la_{1+}(y)))}{\pi (u-v)} +
       O\left(\frac{1}{n}\right).
\end{aligned}
\end{equation}
Because of (\ref{pf2b}) we have by the mean value theorem,
\begin{equation} \label{pf6}
    \Im(\la_{1+}(x) - \la_{1+}(y)) = (x-y) \pi \rho(t)
\end{equation}
for some $t$ between $x$ and $y$. Using (\ref{pf4}) we get $t = x_0 + O(1/n)$ and
\begin{equation} \label{pf7}
    n \Im(\la_{1+}(x) - \la_{1+}(y)) = \pi(u-v) \frac{\rho(t)}{\rho(x_0)} =
    \pi(u-v)\left(1+O\left(\frac{1}{n}\right)\right).
\end{equation}
Inserting (\ref{pf7}) into (\ref{pf5}), we obtain
\begin{equation} \label{pf8}
    \frac{1}{n\rho(x_0)} \hat{K}_n(x,y)
    =  \frac{\sin \pi (u-v)}{\pi (u-v)} + O\left(\frac{1}{n}\right)
\end{equation}
which proves Theorem \ref{maintheo2}.
\hfill \qedsymbol

\subsection{Proof of Theorem \ref{maintheo3}}
We only give the proof of (\ref{m18}), since the proof of
(\ref{m19}) is similar.
We take $\rho_1$ as in (\ref{rhoconst}) and we recall that
$\beta'(z_1) = \rho_1^{2/3}$.

Take $u,v \in \mathbb R$ and let
\begin{equation} \label{pf9}
x = z_1 + \frac{u}{(\rho_1n)^{2/3}}, \qquad y = z_1 + \frac{v}{(\rho_1n)^{2/3}}.
\end{equation}
Assume $u, v < 0$ so that we can use formula (\ref{st14}) for $K_n(x,y)$.
Then we have that $x$ belongs to $D(z_1,r)$, for $n$ large enough, so that
by (\ref{tt1}), (\ref{tildeP}) and (\ref{ep7})
\begin{equation} \label{pf10}
\begin{aligned}
S_+(x) &= R(x) P_+(x) = R(x)\tilde{P}_+(x) \\
    &= R(x) E_n(x) \Phi_+(n^{2/3} \beta(x)) \diag\left(e^{\frac{1}{2}n(\la_1-\la_2)_+(x)}, e^{-\frac{1}{2}n(\la_1-\la_2)_+(x)}, 1\right) \\
    &= R(x) E_n(x) \Phi_+(n^{2/3} \beta(x)) \diag\left(e^{n i \Im \la_{1+}(x)}, e^{-n i \Im \la_{1+}(x)}, 1\right)
\end{aligned}
\end{equation}
and similarly  for $S_+(y)$. Then we get from (\ref{st14}) and (\ref{m16})
\begin{equation} \label{pf11}
\begin{aligned}
\frac{1}{(\rho_1 n)^{2/3}} \hat{K}_n(x,y) &= \frac{1}{2\pi i(u-v)} \begin{pmatrix} -1 & 1 & 0 \end{pmatrix} \Phi_+^{-1}(n^{2/3} \beta(y)) E_n^{-1}(y) R^{-1}(y) \\
 & \quad \times R(x) E_n(x)
    \Phi_+(n^{2/3} \beta(x)) \begin{pmatrix} 1 \\ 1 \\ 0 \end{pmatrix}
    \end{aligned}
\end{equation}

Since $\rho^{2/3} = \beta'(z_1)$, we have as $n \to \infty$,
\begin{equation} \label{pf12}
   n^{2/3} \beta(x) = n^{2/3} \beta\left(z_1 + \frac{u}{(\rho_1n)^{2/3}}\right) \to u
\end{equation}
which implies that $\Phi_+(n^{2/3} \beta(x)) \to \Phi_+(u)$. We use the
second formula of (\ref{ep5}) to evaluate $\Phi_+(u)$ (since $u < 0$), and
it follows that
\begin{equation} \label{pf13}
\begin{aligned}
\lim_{n\to\infty} \Phi_+(n^{2/3} \beta(x)) \begin{pmatrix} 1 \\ 1 \\ 0 \end{pmatrix}
    & = \begin{pmatrix} -y_1(u) - y_2(u) \\ -y_1'(u) - y_2'(u) \\ 0 \end{pmatrix}
    & = \begin{pmatrix} y_0(u) \\ y_0'(u) \\ 0 \end{pmatrix}.
\end{aligned}
\end{equation}
Similarly
\begin{equation} \label{pf14}
\begin{aligned}
    \lim_{n\to\infty} \begin{pmatrix} -1 & 1 & 0 \end{pmatrix}
    \Phi_+^{-1}(n^{2/3} \beta(y))
    & = \begin{pmatrix} - 1 & 1 & 0 \end{pmatrix}
    \Phi_+^{-1}(v) \\
    & =  -2\pi i \begin{pmatrix} - 1 & 1 & 0 \end{pmatrix}
    \begin{pmatrix} -y_2'(v) & y_2(v) & 0 \\ y_1'(v) & -y_1(v) & 0 \\ 0 & 0 & 1 \end{pmatrix} \\
    & = -2\pi i \begin{pmatrix} y_2'(v) + y_1'(v) & -y_2(v) - y_1(v) & 0 \end{pmatrix} \\
    & = -2\pi i \begin{pmatrix} -y_0'(v) & y_0(v) & 0 \end{pmatrix}.
\end{aligned}
\end{equation}
The factor $-2\pi i$ comes from the inverse of $\Phi_+(v)$, since $\det \Phi = (-2\pi i)^{-1}$
by Wronskian relations.

Next, we recall that $R^{-1}(y)R(x) = I + O\left(\frac{x-y}{n}\right)$,
so that by (\ref{pf9})
\begin{equation} \label{pf15}
    R^{-1}(y) R(x) = I + O\left(\frac{1}{n^{5/3}}\right).
\end{equation}
The explicit form (\ref{ep8}) for $E_n$ readily gives
\begin{equation} \label{pf16}
    E_n(x) = O(n^{1/6}), \qquad E_n^{-1}(y) = O(n^{1/6}), \qquad
    E_n^{-1}(y) E_n(x) = I + O\left(\frac{1}{n^{1/3}}\right).
\end{equation}
Combining (\ref{pf15}) and (\ref{pf16}), we have
\begin{equation} \label{pf17}
   \lim_{n \to \infty} E_n^{-1}(y) R^{-1}(y) R(x) E_n(x) = I.
\end{equation}

Inserting (\ref{pf13}), (\ref{pf14}), and (\ref{pf17}) into (\ref{pf11}), we
obtain
\begin{equation} \label{pf18}
\begin{aligned}
\lim_{n \to \infty} \frac{1}{cn^{2/3}} \hat{K}_n(x,y) &= \frac{1}{2\pi i(u-v)}
     \times (-2\pi i) \begin{pmatrix} -y_0'(v) & y_0(v) & 0 \end{pmatrix}
     \begin{pmatrix} y_0(u) \\ y_0'(u) \\ 0 \end{pmatrix}\\
     &= \frac{y_0(u)y_0'(v) - y_0'(u)y_0(v)}{u-v}.
    \end{aligned}
    \end{equation}
Since $y_0 = \Ai$,
we have now completed the proof of (\ref{m18}) in case $u, v < 0$.

\bigskip

For the remaining cases where $u \geq 0$ and/or $v \geq 0$, we have to realize
that we have not specified the rescaled kernel $\hat{K}_n(x,y)$
for $x$ and/or $y$ outside of $[-z_1,-z_2] \cup [z_2,z_1]$, since in (\ref{m20})
$h$ is only defined there. We define
\begin{equation} \label{pf19}
    h(x) = - \frac{1}{4} x^2 + \frac{1}{2} \left(\la_1(x) + \la_2(x)\right),
    \qquad x \in (z_1, \infty).
\end{equation}

We will assume in the rest of the proof that $u > 0$ and $v > 0$.
The case where $u$ and $v$ have opposite signs follows in a similar way:
then we have to combine the calculations given below with
the ones given above.

So let $u, v > 0$ and let $x$ and $y$ be as in (\ref{pf9}).
For the kernel $K_n$ we start from the expression (\ref{ft16}) in terms of $T$.
Since $u > 0$, we have $x, y > z_1$, and so we have by (\ref{st6}), (\ref{tt1}), (\ref{tildeP}) and (\ref{ep7}),
\begin{equation} \label{pf20}
\begin{aligned}
T_+(x) & = S_+(x) = R(x) P_+(x) = R(x) \tilde{P}_+(x)
        \begin{pmatrix} 1 & 0 & 0 \\
        0 & 1 & e^{n(\la_3(x)-\la_2(x))} \\
        0 & 0 & 1
        \end{pmatrix} \\
    & = R(x) E_n(x) \Phi_+(n^{2/3} \beta(x)) \\
    & \qquad \times
        \begin{pmatrix} e^{\frac{1}{2}n(\la_1(x)-\la_2(x))} & 0 & 0 \\
        0 & e^{-\frac{1}{2}n(\la_1(x)-\la_2(x))} & 0 \\
        0 & 0 & 1 \end{pmatrix}
        \begin{pmatrix} 1 & 0 & 0 \\
        0 & 1 & e^{n(\la_3(x)-\la_2(x))} \\
        0 & 0 & 1
        \end{pmatrix}.
\end{aligned}
\end{equation}
Then
\begin{equation} \label{pf21}
T_+(x) \begin{pmatrix} e^{-n \la_1(x)} \\ 0 \\ 0 \end{pmatrix}
    = e^{-\frac{1}{2}n(\la_1(x)- \la_2(x))}
    R(x) E_n(x) \Phi_+(n^{2/3} \beta(x))
    \begin{pmatrix} 1 \\ 0 \\ 0 \end{pmatrix}.
\end{equation}
As before, we have $\Phi_+(n^{2/3} \beta(x)) \to \Phi_+(u)$
as $n \to \infty$. Now we use the first formula of (\ref{ep5})
to evaluate $\Phi_+(u)$ so that
\begin{equation} \label{pf22}
    \lim_{n \to \infty} \Phi_+(n^{2/3} \beta(x) \begin{pmatrix} 1 \\ 0 \\ 0 \end{pmatrix}
    = \Phi_+(u) \begin{pmatrix} 1 \\ 0 \\ 0 \end{pmatrix}
    = \begin{pmatrix} y_0(u) \\ y_0'(u) \\ 0 \end{pmatrix}.
\end{equation}

We have as in (\ref{pf20})
\begin{equation} \label{pf23}
\begin{aligned}
    T_+^{-1}(y) &= \begin{pmatrix} 1 & 0 & 0 \\
        0 & 1 & -e^{n(\la_3(y)-\la_2(y))} \\
        0 & 0 & 1
        \end{pmatrix}
         \begin{pmatrix} e^{-\frac{1}{2}n(\la_1(y)-\la_2(y))} & 0 & 0 \\
        0 & e^{\frac{1}{2}n(\la_1(y)-\la_2(y))} & 0 \\
        0 & 0 & 1 \end{pmatrix} \\
        & \qquad \times
        \Phi_+^{-1}(n^{2/3} \beta(y)) E_n^{-1}(y) R^{-1}(y),
\end{aligned}
\end{equation}
so that
\begin{equation} \label{pf24}
\begin{pmatrix} 0  & e^{n\la_2(y)} & e^{n\la_3(y)} \end{pmatrix}
    T_+^{-1}(y) =
    e^{\frac{1}{2}n(\la_1(y)-\la_2(y))}
    \begin{pmatrix} 0 & 1 & 0 \end{pmatrix} \Phi_+^{-1}(n^{2/3} \beta(y)) E_n^{-1}(y) R^{-1}(y).
\end{equation}
We have
\begin{equation} \label{pf25}
\begin{aligned}
\lim_{n\to\infty} \begin{pmatrix} 0 & 1 & 0 \end{pmatrix} \Phi_+^{-1}(n^{2/3} \beta(y))
& = \begin{pmatrix} 0 & 1 & 0 \end{pmatrix} \Phi_+^{-1}(u) \\
& = (-2\pi i) \begin{pmatrix} -y_0'(u) & y_0(u) & 0 \end{pmatrix}
\end{aligned}
\end{equation}
and as before we have (\ref{pf17}).

Inserting (\ref{pf21}) and (\ref{pf24}) into (\ref{pf16}) and using the
limits (\ref{pf17}), (\ref{pf22}) and (\ref{pf25}), we arrive at
(\ref{m18}) in the case $u, v > 0$.

This completes the proof of Theorem \ref{maintheo3}.
\hfill \qedsymbol

\section{Large $n$ asymptotics of the multiple
Hermite polynomials}\label{sec10}

As noted in Section \ref{sec2}, see also \cite{BK}, we have that the
$(1,1)$ entry of the solution $Y$ of the RH problem (\ref{m4})-(\ref{m5})
is a monic polynomial $P_n$ of degree $n$ satisfying
\[ \int_{-\infty}^{\infty} P_n(x) x^k w_j(x) dx
    = 0, \qquad k=0, 1, \ldots, n_j-1, \qquad j=1,2. \]
For the case $w_1(x) = e^{-n(\frac{1}{2}x^2-ax)}$,
$w_2(x) = e^{-n(\frac{1}{2}x^2+ax)}$, this polynomial is called
a multiple Hermite polynomial \cite{ABVA,VAC}. The asymptotic
analysis of the RH problem  done in Sections \ref{sec4}--\ref{sec9},
also yields the strong asymptotics of the
multiple Hermite polynomials (as $n \to \infty$ with $n$ even and $n_1=n_2$)
in every part of the complex plane. We describe these asymptotics here.
Recall that $P_n$ is the average characteristic polynomial of the
random matrix ensemble (\ref{m1}), see (\ref{Y11}).

We will partition the complex plane into  3 regions:
\begin{enumerate}
\item Outside of the lenses and of the disks $D(\pm z_j,r)$, $j=1,2$.
\item Inside of the lenses but outside of the disks.
\item Inside of the disks.
\end{enumerate}
We will derive the large $n$ asymptotics of the multiple Hermite polynomials
in these 3 regions.

\subsection*{(1) Region  outside of the lenses and of the disks}
In this region, we have by (\ref{st6}) and (\ref{tt1}),
\begin{equation} \label{l1}
T(z)=R(z)M(z),
\end{equation}
hence by (\ref{defT})
\begin{equation} \label{l2}
\begin{aligned}
\diag(e^{-nl_1},e^{-nl_2},e^{-nl_3})&Y(z)
\diag(e^{-\frac{n}{2}z^2},e^{-naz},e^{naz})\\
{}&=R(z)M(z)
\diag(e^{-n\la_1(z)},e^{-n\la_2(z)},e^{-n\la_3(z)}).
\end{aligned}
\end{equation}
By restricting this matrix equation to the element $(1,1)$
we obtain that
\begin{equation} \label{l3}
P_n(z)e^{-\frac{n}{2}z^2}=e^{-n\la(z)}\sum_{j=1}^3R_{1j}(z)M_{j1}(z),
\end{equation}
where
\begin{equation} \label{l4}
\la(z)\equiv \la_1(z)-l_1=\int^z\xi_1(s)\,ds,
\end{equation}
and as $z\to\infty$,
\begin{equation} \label{l5}
\la(z)=\frac{z^2}{2}-\ln z+O(z^{-2}).
\end{equation}
In the sum over $j$ in (\ref{l3}) the term $j=1$ dominates
and we obtain because of (\ref{mod17}) that
\begin{equation} \label{l6}
P_n(z)e^{-\frac{n}{2}z^2}=
\frac{\xi_1^2(z)-a^2}{\sqrt{(\xi_1^2(z)-p^2)(\xi_1^2(z)-q^2)}}
e^{-n\la(z)}\left(1+O\left(\frac{1}{n(|z|+1)}\right)\right),
\end{equation}
where for the square root we use the principal branch
(the one that is positive for $z>z_1$), with two cuts,
$[-z_1,-z_2]$ and $[z_2,z_1]$.

\subsection*{(2) Region  inside of the lenses but outside
of the disks.}
In this region, we get from (\ref{st2}), (\ref{st4}) and (\ref{tt1}),
\begin{equation} \label{l7}
T(z)=R(z)M(z)L(z)^{-1},
\end{equation}
where $L(z)$ is the matrix on the right in
(\ref{st2}) and (\ref{st4}). Hence by (\ref{defT})
\begin{equation} \label{l8}
\begin{aligned}
\diag(e^{-nl_1},e^{-nl_2},e^{-nl_3})&Y(z)
\diag(e^{-\frac{n}{2}z^2},e^{-naz},e^{naz})\\
{}&=R(z)M(z)L(z)^{-1}
\diag(e^{-n\la_1(z)},e^{-n\la_2(z)},e^{-n\la_3(z)}).
\end{aligned}
\end{equation}
Consider $z$ the upper lens region on $[z_2,z_1]$.
Then
\begin{equation} \label{l10}
L(z)=\begin{pmatrix}
1 & 0  & 0 \\
-e^{n(\la_1(z)-\la_2(z))} & 1 & -e^{n(\la_3(z)-\la_2(z))} \\
0 & 0 & 1
\end{pmatrix},
\end{equation}
hence
\begin{equation} \label{l11}
L(z)^{-1}=\begin{pmatrix}
1 & 0  & 0 \\
e^{n(\la_1(z)-\la_2(z))} & 1 & e^{n(\la_3(z)-\la_2(z))} \\
0 & 0 & 1
\end{pmatrix},
\end{equation}
and the first column of the
matrix $M(z)L(z)^{-1}
\diag(e^{-n\la_1(z)},e^{-n\la_2(z)},e^{-n\la_3(z)})$ is
\begin{equation} \label{l12}
\begin{pmatrix}
M_1(\xi_1(z))e^{-n\la_1(z)}+M_1(\xi_2(z))e^{-n\la_2(z)}  \\
M_2(\xi_1(z))e^{-n\la_1(z)}+M_2(\xi_2(z))e^{-n\la_2(z)}  \\
M_3(\xi_1(z))e^{-n\la_1(z)}+M_3(\xi_2(z))e^{-n\la_2(z)}
\end{pmatrix},
\end{equation}
see (\ref{mod7}).
By restricting equation (\ref{l8}) to the $(1,1)$ entry,
and using (\ref{mod17}) and (\ref{tt6}),
we obtain that in the upper lens region on $[z_2,z_1]$
\begin{equation} \label{l13}
\begin{aligned}
P_n(z)e^{-\frac{n}{2}z^2}&=
\left[\frac{\xi_1^2(z)-a^2}{\sqrt{(\xi_1^2(z)-p^2)(\xi_1^2(z)-q^2)}}
+O\left(\frac{1}{n}\right)\right]
e^{-n\la_1(z)+nl_1}\\
{}& \qquad +\left[\frac{\xi_2^2(z)-a^2}{\sqrt{(\xi_2^2(z)-p^2)(\xi_2^2(z)-q^2)}}
+O\left(\frac{1}{n}\right)\right]
e^{-n\la_2(z)+nl_1},
\end{aligned}
\end{equation}
where
\begin{equation} \label{l14}
\la_k(z)=\int_{z_1}^z\xi_k(s)\,ds,\qquad k=1,2.
\end{equation}
In the same way we obtain that in the lower lens region on
$[z_2,z_1]$,
\begin{equation} \label{l13a}
\begin{aligned}
P_n(z)e^{-\frac{n}{2}z^2}&=
\left[\frac{\xi_1^2(z)-a^2}{\sqrt{(\xi_1^2(z)-p^2)(\xi_1^2(z)-q^2)}}
+O\left(\frac{1}{n}\right)\right]
e^{-n\la_1(z)+nl_1}\\
{}& \qquad -\left[\frac{\xi_2^2(z)-a^2}{\sqrt{(\xi_2^2(z)-p^2)(\xi_2^2(z)-q^2)}}
+O\left(\frac{1}{n}\right)\right]
e^{-n\la_2(z)+nl_1}.
\end{aligned}
\end{equation}
For $z=x$ real, $x\in[z_2+r,z_1-r]$, both (\ref{l13}) and (\ref{l13a})
 can be rewritten in the form
\begin{equation} \label{l15}
P_n(x)e^{-\frac{n}{2}x^2}=\left\{A(x)\cos[n\,\Im \la_{1+}(x)-\f(x)]
+O\left(\frac{1}{n}\right)\right\}e^{-n\,\Re\la_{1+}(x)+nl_1},
\end{equation}
where
\begin{equation} \label{l16}
A(x)=2\left|\frac{\xi_{1+}^2(x)-a^2}
{\sqrt{(\xi_{1+}^2(x)-p^2)(\xi_{1+}^2(x)-q^2)}}\right|
\end{equation}
and
\begin{equation} \label{l17}
\f(x)=\arg\frac{\xi_{1+}^2(x)-a^2}
{\sqrt{(\xi_{1+}^2(x)-p^2)(\xi_{1+}^2(x)-q^2)}}\,.
\end{equation}
By using equation (\ref{defrho}), we can also rewrite
(\ref{l15}) in terms of the eigenvalue density function $\rho(x)$,
\begin{equation} \label{l15a}
P_n(x)e^{-\frac{n}{2}x^2}=\left\{A(x)\cos\left[n\pi\int_{z_1}^x
\rho(s)\,ds-\f(x)\right]
+O\left(\frac{1}{n}\right)\right\}e^{-n\,\Re \la_{1+}(x)+nl_1}.
\end{equation}
Equation (\ref{l15a}) clearly displays the oscillating behavior of
$P_n$ on the interval $[z_2+r, z_1-r]$. It also shows that the zeros
of $P_n(x)$ are asymptotically distributed like $\rho(x)dx$, the
limiting probability distribution of eigenvalues.
Similar formulae can be derived on the interval $[-z_1+r,-z_2-r]$.

\subsection*{(3) Region inside of the disks.} Consider the disk
$D(z_1,r)$. In the regions I and IV, we have by
(\ref{tildeP}), (\ref{tt1}) and (\ref{tt6})
\begin{equation} \label{l19}
T(z)=R(z)P(z)=\left(I+O(n^{-1})\right)\tilde P(z)
\begin{pmatrix}
1 & 0 & 0 \\
0 & 1 & e^{n(\la_3(z)-\la_2(z))} \\
0 & 0 & 1
\end{pmatrix},
\end{equation}
hence by (\ref{defT}), (\ref{ep7}), and (\ref{ep8})
\begin{equation} \label{l20}
\begin{aligned}
\diag&(e^{-nl_1},e^{-nl_2},e^{-nl_3})Y(z)
\diag(e^{-\frac{n}{2}z^2},e^{-naz},e^{naz})\\
{}&=(I+O(n^{-1}))\sqrt\pi M(z)
\begin{pmatrix}
n^{1/6}\be(z)^{1/4} & -n^{-1/6}\be(z)^{-1/4} & 0 \\
-in^{1/6}\be(z)^{1/4} & -in^{-1/6}\be(z)^{-1/4} & 0 \\
0 & 0 & 1
\end{pmatrix}\\
{}& \qquad \times\Phi(n^{2/3}\be(z))
\diag(e^{-n\al(z)},e^{-n\al(z)},e^{-n\la_3(z)})
\begin{pmatrix}
1 & 0 & 0 \\
0 & 1 & 1 \\
0 & 0 & 1
\end{pmatrix},
\end{aligned}
\end{equation}
where
\begin{equation} \label{l21}
\al(z)=\frac{\la_1(z)+\la_2(z)}{2}\,.
\end{equation}
By restricting equation (\ref{l20}) to the $(1,1)$ entry,
and using the first expression of (\ref{ep5}) (in region I)
or the fourth expression of (\ref{ep5}) (in region IV) to
evaluate $\Phi(n^{2/3} \be(z))$, and (\ref{mod17}) to evaluate
$M(z)$, we obtain that
\begin{equation} \label{l22}
\begin{aligned}
P_n(z)e^{-\frac{n}{2}z^2}&=\sqrt\pi\,
\left[\,n^{1/6}B(z)\Ai(n^{2/3}\be(z))(1+O(n^{-1})) \right. \\[10pt]
{}& \qquad \qquad
\left. +n^{-1/6}C(z)\Ai'(n^{2/3}\be(z))(1+O(n^{-1}))\, \right]\, 
e^{-n\al(z)+nl_1},
\end{aligned}
\end{equation}
where
\begin{equation} \label{l23}
B(z)=\be(z)^{1/4}\left(
\frac{\xi_1^2(z)-a^2}
{\sqrt{(\xi_1^2(z)-p^2)(\xi_1^2(z)-q^2)}}
-i\frac{\xi_2^2(z)-a^2}
{\sqrt{(\xi_2^2(z)-p^2)(\xi_2^2(z)-q^2)}}\right)
\end{equation}
and
\begin{equation} \label{l24}
C(z)=\be(z)^{-1/4}\left(
-\frac{\xi_1^2(z)-a^2}
{\sqrt{(\xi_1^2(z)-p^2)(\xi_1^2(z)-q^2)}}
-i\frac{\xi_2^2(z)-a^2}
{\sqrt{(\xi_2^2(z)-p^2)(\xi_2^2(z)-q^2)}}\right).
\end{equation}
The same asymptotics, (\ref{l22}), holds in regions II and III as well.
Thus, (\ref{l22}) holds in the full disk $D(z_1,r)$.
It may be verified that the functions $B(z)$ and $C(z)$ are analytic in
$D(z_1,r)$.

This approach allows one to derive
a formula similar to (\ref{l22}) in all the other disks
$D(\pm z_j,r)$ as well.

\appendix
\section{Recurrence equations for multiple Hermite polynomials} \label{REC}

From orthogonality equation
(\ref{m8}), we obtain that as $z\to\infty$,
\begin{equation}\label{rec1}
\begin{aligned}
\frac{1}{2\pi i}\int_{-\infty}^\infty \frac{P_{n_1,n_2}(u)w_k(u)}
{u-z}\,du&=-\frac{1}{2\pi i}\int_{-\infty}^\infty P_{n_1,n_2}(u)w_k(u)
\left(\frac{1}{z}+\frac{u}{z^2}+\cdots\right)\,du\\
&=-\frac{1}{2\pi i}\left(\frac{h^{(k)}_{n_1,n_2}}{z^{n_k+1}}+
\frac{q^{(k)}_{n_1,n_2}}{z^{n_k+2}}+\cdots\right),\quad k=1,2,
\end{aligned}
\end{equation}
where for $k=1,2$, $h^{(k)}_{n_1,n_2}$ is defined in (\ref{m9}) and
\begin{equation}\label{rec2}
q^{(k)}_{n_1,n_2}=\int_{-\infty}^\infty
P_{n_1,n_2}(x)x^{n_k+1}w_k(x)dx.
\end{equation}
This implies that
\begin{equation}\label{rec3}
\begin{aligned}
\Psi_{n_1,n_2}(z)&=\left(I+\frac{\Psi^{(1)}_{n_1,n_2}}{z}
+\cdots\right)
\diag\left(z^n e^{-\frac{1}{2}Nz^2},c_1^{-1}z^{-n_1} e^{-Naz},
c_2^{-1}z^{-n_2}e^{Naz}\right)
\end{aligned}
\end{equation}
where
\begin{equation}\label{rec4}
\Psi^{(1)}_{n_1,n_2}=
\begin{pmatrix}
p_{n_1,n_2} & \frac{h^{(1)}_{n_1,n_2}}{h^{(1)}_{n_1-1,n_2}} &
\frac{h^{(2)}_{n_1,n_2}}{h^{(2)}_{n_1,n_2-1}}  \\
1 & \frac{q^{(1)}_{n_1-1,n_2}}{h^{(1)}_{n_1-1,n_2}}
& \frac{h^{(2)}_{n_1-1,n_2}}{h^{(2)}_{n_1,n_2-1}} \\
1 & \frac{h^{(1)}_{n_1,n_2-1}}{h^{(1)}_{n_1-1,n_2}}
& \frac{q^{(2)}_{n_1,n_2-1}}{h^{(2)}_{n_1,n_2-1}}
\end{pmatrix},
\end{equation}
and $P_{n_1,n_2}(z)=z^n+p_{n_1,n_2}z^{n-1}+\cdots$.
Set
\begin{equation}\label{rec5}
U_{n_1,n_2}(z)=\Psi_{n_1+1,n_2}(z)\Psi_{n_1,n_2}(z)^{-1}.
\end{equation}
Then by (\ref{m15}), $U_{n_1,n_2+}(x)=U_{n_1,n_2-}(x)$
(i.e., no jump on the real line) and as $z\to\infty$,
\begin{equation}\label{rec6}
\begin{aligned}
U_{n_1,n_2}(z)&\cong\left(I+\frac{\Psi^{(1)}_{n_1+1,n_2}}{z}
+\cdots\right)
\begin{pmatrix}
z & 0 & 0 \\
0 & z^{-1}\frac{h^{(1)}_{n_1,n_2}}{h^{(1)}_{n_1-1,n_2}} & 0 \\
0 & 0 & \frac{h^{(2)}_{n_1+1,n_2-1}}{h^{(2)}_{n_1,n_2-1}}
\end{pmatrix}
\left(I+\frac{\Psi^{(1)}_{n_1,n_2}}{z}
+\cdots\right)^{-1}\\
&=zP_1+\Psi^{(1)}_{n_1+1,n_2}P_1-P_1\Psi^{(1)}_{n_1,n_2}
+\frac{h^{(2)}_{n_1+1,n_2-1}}{h^{(2)}_{n_1,n_2-1}}P_3
+O\left(\frac{1}{z}\right),
\end{aligned}
\end{equation}
where
\begin{equation}\label{rec7}
P_1=\diag(1,0,0),
\quad
P_2=\diag(0,1,0),
\quad
P_3=\diag(0,0,1).
\end{equation}
Since $U_{n_1,n_2}(z)$ is analytic on the complex plane, equation
(\ref{rec6}) implies, by the Liouville theorem, that
\begin{equation}\label{rec8}
\begin{aligned}
U_{n_1,n_2}(z)&=zP_1+\Psi^{(1)}_{n_1+1,n_2}P_1-P_1\Psi^{(1)}_{n_1,n_2}
+\frac{h^{(2)}_{n_1+1,n_2-1}}{h^{(2)}_{n_1,n_2-1}}P_3\\
{}&=
\begin{pmatrix}
z-b_{n_1,n_2} & -c_{n_1,n_2} & -d_{n_1,n_2}  \\
1 & 0 & 0 \\
1 & 0 & e_{n_1,n_2}
\end{pmatrix},
\end{aligned}
\end{equation}
where
\begin{equation}\label{rec9}
c_{n_1,n_2}
=\frac{h^{(1)}_{n_1,n_2}}{h^{(1)}_{n_1-1,n_2}}\not=0,
\quad
d_{n_1,n_2}
=\frac{h^{(2)}_{n_1,n_2}}{h^{(2)}_{n_1,n_2-1}}\not=0,\quad
e_{n_1,n_2}=\frac{h^{(2)}_{n_1+1,n_2-1}}{h^{(2)}_{n_1,n_2-1}}\not=0.
\end{equation}
Thus, we obtain the matrix recurrence equation,
\begin{equation}\label{rec10}
\Psi_{n_1+1,n_2}(z)=U_{n_1,n_2}(z)\Psi_{n_1,n_2}(z).
\end{equation}
By restricting it to the element $(1,1)$ we obtain that
\begin{equation}\label{rec11}
P_{n_1+1,n_2}(z)=(z-b_{n_1,n_2})P_{n_1,n_2}(z)
-c_{n_1,n_2}P_{n_1-1,n_2}(z)
-d_{n_1,n_2}P_{n_1,n_2-1}(z),
\end{equation}
and by restricting it to the element $(3,1)$ we obtain that
\begin{equation}\label{rec12}
P_{n_1+1,n_2-1}(z)=P_{n_1,n_2}(z)
+e_{n_1,n_2}P_{n_1,n_2-1}(z).
\end{equation}
Similar to (\ref{rec10}), we have another recurrence equation,
\begin{equation}\label{rec13}
\Psi_{n_1,n_2+1}(z)=\tilde U_{n_1,n_2}(z)\Psi_{n_1,n_2}(z),
\end{equation}
where
\begin{equation}\label{rec14}
\tilde U_{n_1,n_2}(z)=\begin{pmatrix}
z-\tilde b_{n_1,n_2} & - c_{n_1,n_2}
& - d_{n_1,n_2}  \\
1 & \tilde e_{n_1,n_2} & 0 \\
1 & 0 & 0
\end{pmatrix},
\end{equation}
and
\begin{equation}\label{rec15}
\tilde e_{n_1,n_2}
=\frac{h^{(1)}_{n_1-1,n_2+1}}{h^{(1)}_{n_1-1,n_2}}\not=0.
\end{equation}
By restricting (\ref{rec13}) to the elements
$(1,1)$ and $(2,1)$, we obtain the equations,
\begin{equation}\label{rec16}
P_{n_1,n_2+1}(z)=(z-\tilde b_{n_1,n_2})P_{n_1,n_2}(z)
-c_{n_1,n_2}P_{n_1-1,n_2}(z)
-d_{n_1,n_2}P_{n_1,n_2-1}(z),
\end{equation}
and
\begin{equation}\label{rec17}
P_{n_1-1,n_2+1}(z)=P_{n_1,n_2}(z)
+\tilde e_{n_1,n_2}P_{n_1-1,n_2}(z).
\end{equation}

\section{Differential equations for multiple Hermite polynomials} \label{DIF}

Set
\begin{equation}\label{de1}
A_{n_1,n_2}(z)=\frac{1}{N}\,\Psi_{n_1,n_2}'(z)\Psi_{n_1,n_2}(z)^{-1}.
\end{equation}
It follows from (\ref{m15}), that $A_{n_1,n_2}(z)$ has no jump
on the real axis, so that it is analytic on the complex plane.
By differentiating (\ref{rec3}) we obtain that  as $z\to\infty$,
\begin{equation}\label{de3}
A_{n_1,n_2}(z)=\left(I+\frac{\Psi^{(1)}_{n_1,n_2}}{z}+\cdots\right)
\begin{pmatrix}
-z & 0 & 0 \\
0 & -a  & 0 \\
0 & 0 & a
\end{pmatrix}
\left(I+\frac{\Psi^{(1)}_{n_1,n_2}}{z}+\cdots\right)^{-1}
+O\left(\frac{1}{z}\right).
\end{equation}
Since $A_{n_1,n_2}(z)$ is analytic, we obtain that
\begin{equation}\label{de4}
\begin{aligned}
A_{n_1,n_2}(z)&=-\left[\left(I+\frac{\Psi^{(1)}_{n_1,n_2}}{z}
+\cdots\right)
\begin{pmatrix}
z & 0 & 0 \\
0 & 0 & 0 \\
0 & 0 & 0
\end{pmatrix}
\left(I+\frac{\Psi^{(1)}_{n_1,n_2}}{z}+\cdots\right)^{-1}\right]_{\rm pol}\\
{}&+\begin{pmatrix}
0 & 0 & 0 \\
0 & -a & 0 \\
0 & 0 & a
\end{pmatrix},
\end{aligned}
\end{equation}
where $[f(z)]_{\rm pol}$ means the polynomial part of $f(z)$ at
infinity.  From (\ref{de1}) we get the differential equation,
\begin{equation}\label{de5}
\Psi_{n_1,n_2}'(z)=N A_{n_1,n_2}(z)\Psi_{n_1,n_2}(z).
\end{equation}
and (\ref{de4}) reduces to
\begin{equation}\label{de7}
A_{n_1,n_2}(z)=\begin{pmatrix}
-z
 & c_{n_1,n_2} & d_{n_1,n_2}  \\
{}-1 & -a & 0 \\
{}-1 & 0 & a
\end{pmatrix}.
\end{equation}

\section{Proof of Proposition \ref{Lax}}\label{AI}

Equations (\ref{rec10}), (\ref{rec13}), (\ref{de5}) form a Lax pair for
multiple Hermite polynomials.
Their compatibility conditions are
\begin{equation}\label{se1}
\begin{aligned}
\frac{1}{N}U_{n_1,n_2}'(z) & =A_{n_1+1,n_2}(z)U_{n_1,n_2}(z)
-U_{n_1,n_2}(z)A_{n_1,n_2}(z), \\
\frac{1}{N}\tilde{U}_{n_1,n_2}'(z) & =A_{n_1,n_2+1}(z)\tilde{U}_{n_1,n_2}(z)
-\tilde{U}_{n_1,n_2}(z)A_{n_1,n_2}(z).
\end{aligned}
\end{equation}
This gives the equations,
\begin{equation}\label{se3}
\begin{aligned}
b_{n_1,n_2}=a, \quad
 c_{n_1+1,n_2}=c_{n_1,n_2}+\frac{1}{N}, \quad
& d_{n_1+1,n_2}=d_{n_1,n_2}, \quad \quad
e_{n_1,n_2}=-2a, \\
\tilde{b}_{n_1,n_2} = -a, \quad
 c_{n_1,n_2+1} = c_{n_1,n_2}, \qquad \quad
& d_{n_1,n_2+1} = d_{n_1,n_2} + \frac{1}{N}, \quad
\tilde{e}_{n_1,n_2}  = 2a.
\end{aligned}
\end{equation}
Since $c_{0,n_2}=d_{n_1,0}=0$, we obtain that
\begin{equation}\label{se5}
c_{n_1,n_2}=\frac{n_1}{N},\quad
d_{n_1,n_2}=\frac{n_2}{N}\,.
\end{equation}
This proves the first equation in (\ref{Lax1}) and equation
(\ref{Lax2}). Similarly we obtain that $\tilde e_{n_1,n_2}=2a$
and this proves the second equation in (\ref{Lax1}). Proposition
\ref{Lax} is proved.
\hfill \qedsymbol

\end{document}